\documentclass[]{aa}

\usepackage[utf8]{inputenc}
\usepackage{natbib}
\usepackage{graphicx}
\usepackage{subcaption}
\usepackage{numprint}
\usepackage{color}
\usepackage{txfonts}
\usepackage{hyperref}
\usepackage{tabularx}
\usepackage{amsmath}
\usepackage{amssymb}
\usepackage{mathtools}
\usepackage[normalem]{ulem}
\usepackage{tikz}
\usetikzlibrary{calc,shapes,arrows}
\usepackage{multirow}
\usepackage{threeparttable}
\providecommand{\Unit}[1]{\ensuremath{\mathrm{~#1}}} 

\providecommand{\parallax}{\ensuremath{\varpi}}

\newcommand\gaia{\textit{Gaia~}}
\newcommand\hip{\textsc{Hipparcos~}}

\newcommand\tyctwo{\textit{Tycho}-2~}
\newcommand\gdrone{\gaia DR1~}
\newcommand\gdrtwo{\gaia DR2~}

\newcommand\secref[1]{Sect.~\ref{#1}}

\newcommand\figref[1]{Fig.~\ref{#1}}

\newcommand\figrefalt[1]{Figure~\ref{#1}}

\newcommand\tabref[1]{Table~\ref{#1}}
\begin{document}

\title{A new method for unveiling Open Clusters in \textit{Gaia}:}
\subtitle{new nearby Open Clusters confirmed by DR2}

\author{
		A. Castro-Ginard			\inst{\ref{inst:UB}}\relax
\and	C. Jordi					\inst{\ref{inst:UB}}\relax
\and	X. Luri						\inst{\ref{inst:UB}}\relax
\and 	F. Julbe					\inst{\ref{inst:UB}}\relax
\and	M. Morvan					\inst{\ref{inst:UB},\ref{inst:MSE}}\relax
\and	L. Balaguer-N\'{u}\~{n}ez	\inst{\ref{inst:UB}}\relax
\and 	T. Cantat-Gaudin			\inst{\ref{inst:UB}}\relax
}

\institute{
    Dept. Fisica Quantica i Astrofisica, Institut de Ciències del Cosmos (ICCUB), Universitat de Barcelona (IEEC-UB), Martí Franquès 1, E08028 Barcelona, Spain\\
    \email{acastro@fqa.ub.edu}\relax \label{inst:UB}
\and
	Mines Saint-Etienne, Institut Henri Fayol, F - 42023 Saint-Etienne France\relax \label{inst:MSE}
}

\date{Received May 8, 2018 /
Accepted June 11, 2018}

\abstract{
 The publication of the \gaia Data Release 2 (\gaia DR2) opens a new era in Astronomy. It includes precise astrometric data (positions, proper motions and parallaxes) for more than $1.3$ billion sources, mostly stars. To analyse such a vast amount of new data, the use of data mining techniques and machine learning algorithms are mandatory.
}{
 The search for Open Clusters, groups of stars that were born and move together, located in the disk, is a great example for the application of these techniques. Our aim is to develop a method to automatically explore the data space, requiring minimal manual intervention.
}{
 We explore the performance of a density based clustering algorithm, DBSCAN, to find clusters in the data together with a supervised learning method such as an Artificial Neural Network (ANN) to automatically distinguish between real Open Clusters and statistical clusters.
}{
The development and implementation of this method to a $5$-Dimensional space ($l$, $b$, $\parallax$, $\mu_{\alpha^*}$, $\mu_\delta$) to the Tycho-Gaia Astrometric Solution (TGAS) data, and a posterior validation using \gdrtwo data, lead to the proposal of a set of new nearby Open Clusters.
}{
We have developed a method to find OCs in astrometric data, designed to be applied to the full \gdrtwo archive.
}
\maketitle


\section{Introduction}
\label{sec:intro}

The volume of data in the astronomical catalogues is continuously increasing with time, and thus its analysis is becoming a highly complex task. In this context, the \gaia mission, with the publication of its first data release \citep[\gaia DR1][]{2016A&A...595A...2G} containing positions for more than one billion sources, opened a new era in Astronomy. But in spite of this large number of stars, full $5$-parameter astrometric data, \textit{i.e.} positions, parallax and proper motions $(\alpha, \delta, \parallax, \mu_{\alpha^*}, \mu_{\delta})$ are available only for a relatively small subset. This subset is the \textit{Tycho-Gaia} Astrometric Solution \citep[TGAS][]{gdr1-tgas,tgas} and it provides a good starting point to devise and test scientific analysis tools in preparation for the larger releases, and in particular for the just published second \gaia data release \citep[\gdrtwo][]{gdr2}. In \gdrtwo precise $5$-parameter astrometric data for more than $1.3$ billion stars are available, together with three band photometry. The analysis of such a vast amount of data is simply not possible with the usual techniques that require a manual supervision, and has to rely on the use of data mining techniques and machine learning algorithms. In this paper we develop a set of such techniques, allowing an automatic exploration of the data space for the detection of open clusters, we apply them to TGAS and we check the validity of the results with the DR2 data, in preparation for its application to the full dataset. 

The analysis tools developed in this paper are designed for the automated detection of open clusters (OC). According to the current accepted scenarios of star formation, most of the stars are born in groups from giant molecular clouds \citep[see for instance][]{EALKMSPCM-1993}. Such groups, of up to few thousand stars, can lose members or even completely dissolve due to internal and close external encounters with stars and gas clouds in their orbits in the Galactic disc. Open clusters being the fundamental building blocks of galaxies are key objects for several astrophysical aspects. To cite some: (a) very young OCs are informative of the star formation mechanism (the fragmentation of the gas clouds, the time sequence of formation, the IMF), (b) young OCs trace the star forming regions (young clusters are seen near their birth place), (c) the evaporation of OCs stars into the field stellar population (by studying the internal kinematics, the mass segregations), (d) intermediate and old OCs allow studying the chemical enrichment of the galactic disc because the more precise determination of ages than for field stars (gradients with galactocentric distance and age can be analysed), (e) the stellar structure and evolution (colour magnitude diagrams CMDs provide empirical isochrones to compare with the theoretical models). The most updated and complete compilations of known OCs are those in \cite{dias} and \cite{kharchenko}\footnote{supplemented by \cite{2014A&A...568A..51S} and \cite{2015A&A...581A..39S}}. Both lists are internally homogeneous in their determination of mean proper motions, distances, reddening and ages, but there is no full agreement between them on which group of stars is considered a cluster or an asterism. In total there are about $2500$ known OCs, most of them detected as stellar overdensities in the sky and confirmed through proper motions and/or CMDs. About $50\%$ of the OCs in these samples are closer than $2$~kpc and about $90\%$ are closer than $5$~kpc. Certainly, our knowledge of OCs beyond $1-2$~kpc is rather incomplete because the decreasing angular size and luminosity of the clusters with distance and because the obscuration by the interstellar dust. \cite{2017MNRAS.469.1545F} identified $125$ compact (distant) and so far unknown OCs using deep high resolution near infrared surveys, again by identifying overdensities in the spatial distribution confirmed as OC by using CMDs. 

The just released \gdrtwo provides an ideal dataset for the detection of so far unknown OCs. Identifying clustering of objects in a multidimensional space (positions, proper motions, parallaxes and photometry) allows for a much more efficient detection of these objects than just in the usual two-dimensional (sky positions) approach. With this purpose in mind we have devised a method to systematically search for OCs in \gaia data in an automatic way and we have, as an initial validation step, applied it to the TGAS subset of \gdrone \citep{2016A&A...595A...2G}. Although the $2$ million stars in TGAS have a relative bright limiting magnitude of $\sim 12$, the inclusion of the proper motions and parallaxes allows us to detect sparse or poorly populated clusters that have been so far gone undetected in the solar neighbourhood\footnote{for instance \cite{SRESBG-2016} discovered nine OCs within $500\Unit{pc}$ from the Sun based of proper motion analysis using a combination of \tyctwo and URAT$1$ catalogues. So, the existence of still undiscovered nearby OCs cannot be discarded.}. Basically, the inclusion of additional dimensions and the better precision of the data increases the statistical significance of the overdensities. These overdensities are detected using a density based clustering algorithm named DBSCAN \citep{dbscan}, which has been previously used to find spatial overdensities \citep{JACLD-2008} or cluster membership determination \citep{XHGLC-2014,XHGCW-2017}. They are subjected to a confirmation step using a classification algorithm based on an Artificial Neural Network \citep{ann} to recognize isochrone patterns on CMDs. The so detected candidate OCs are finally validated by hand using \gdrtwo \citep{gdr2} photometric data, in order to confirm the validity of the methodology in view of its application to the full \gdrtwo archive in an upcoming paper.

This paper is organized as follows: in \secref{sec:method} we describe the clustering algorithm used. In \secref{sec:simu} we optimize the choice of the values of the algorithm parameters by applying it to a simulated dataset. In \secref{sec:identification} the neural network classification algorithm used to discriminate between real OC and detections due to random noise is described. In \secref{sec:results} we discuss the results of the method when applied to the TGAS dataset, materialized in a list of $31$ OC candidates. Finally, these candidates are manually validated using \gdrtwo photometric data in \secref{sec:dr2val}, allowing to confirm most of them. Conclusions are presented in \secref{sec:conclusions}.

\section{Methods}
\label{sec:method}

The methodology used in order to identify groups of stars as possible new OCs is sketched in \figref{fig:flowchart}. Starting from the whole TGAS catalogue and after applying a preprocessing step (see \secref{subsec:preprocessing}), an unsupervised clustering algorithm named DBSCAN\footnote{\label{fn:sklearn}Algorithm from the scikit-learn python package \citep{sklearn}.} detects statistical clusters (see \secref{subsec:dbscan}) in the data. After removing the OCs already catalogued in MWSC, an Artificial Neural Network\textsuperscript{\ref{fn:sklearn}} is applied to automate the distinction between statistical clusters and physical OCs, based on a CMD built using the photometric data from the $2$MASS catalogue.

\tikzstyle{block} = [draw,rectangle,fill=white,rounded corners]
\tikzstyle{cloud} = [draw,ellipse,fill=white]
\tikzstyle{cloud1} = [draw,dashed,ellipse,fill=white]
\tikzstyle{line} = [draw, -latex']
\begin{figure}[htb]
\centering
\begin{tikzpicture}[node distance = 2cm, auto]
\node [cloud1] at (0,1.5) (tgas) {TGAS};
\node [block] at (0,0) (preprocess) {Preprocess};
\node [block] at (0,-1.5) (dbscan) {DBSCAN};
\node [cloud] at (4,-1.5) (found_1) {DBSCAN Clusters};
\node [block] at (4,-3) (ann) {ANN};
\node [cloud] at (4,-4.5) (found_2) {New OCs};
\node [cloud1] at (0,-3) (dr1clust) {CMDs of \gdrone OCs};
\node [cloud1] at (4,0) (kharchenko) {MWSC};

\path [line,dashed] (tgas) -- (preprocess);
\path [line] (preprocess) -- (dbscan);
\path [line,rounded corners] (dbscan) |- ($(dbscan.south west) + (-0.5,-0.5)$) |- (dbscan) node[pos=0.5,sloped,above] {shift};
\path [line] (dbscan) -- (found_1);
\path [line] (found_1) -- (ann);
\path [line] (ann) -- (found_2);
\path [line,dashed] (dr1clust) -- (ann);
\path [line,dashed,rounded corners] (kharchenko.east) -| (6.5,-2.25) -| (4,-2.25);
\end{tikzpicture}
\caption{Flow chart of the method applied to find OCs. Solid boxes represent code, solid ellipses represent generated catalogues while dashed ellipses represent external catalogues.}
\label{fig:flowchart}
\end{figure}
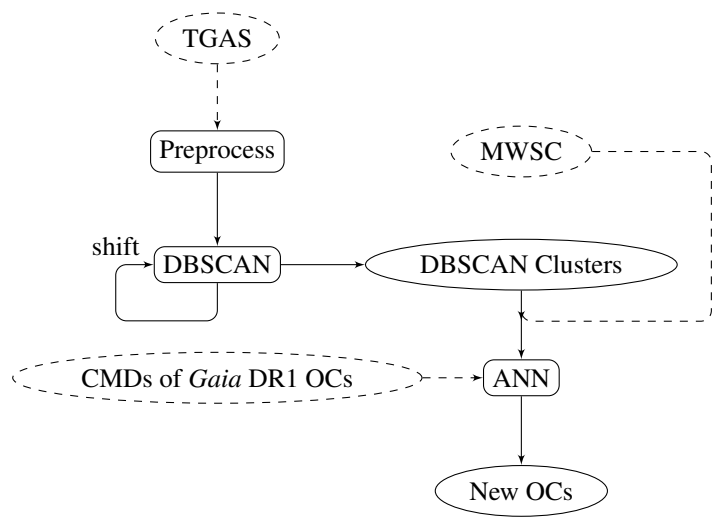

\subsection{Preprocessing}
\label{subsec:preprocessing}

Most of the catalogued Open Clusters are found in the Galactic disk ($|b| < 20 \Unit{deg}$), \textit{e.g.} $96\%$ of the clusters from Dias catalogue \citep{dias} and $94\%$ from the MWSC \citep{kharchenko} lie in that region. So, we explore the Milky Way disk scanning all longitudes in the region $\pm20\Unit{deg}$ in latitude. In addition, we remove stars with extreme proper motions and large or negative parallaxes. This helps in the determination of the DBSCAN parameter $\epsilon$ (see \secref{subsec:dbscan}) with almost no loss of generality because these conditions would make any OC easily detectable. The values to consider a star as rejectable for the algorithm are $|\mu_{\alpha^*}|,|\mu_{\delta}| > 30\Unit{mas\cdot yr}^{-1}$, $\parallax < 0\Unit{mas}$ and $\parallax > 7\Unit{mas}$.

The resulting sky area of study is further divided into smaller regions, rectangles of size $L\Unit{deg}$, where the clustering algorithm is going to be applied. The reason of this division is twofold. On one hand, it saves computational time because the volume of the data in the region is much smaller. On the other hand, the DBSCAN algorithm needs a starting point to define an averaged density of stars in the region; with smaller regions this average is more representative than if we take the whole sky, where the density can significantly vary from one region to another. Once we have the sky divided in rectangles, to avoid the redundant detection of split clusters that might be spread in more than one of these regions or may be in the intersection of two regions, any cluster found with at least one star on the edge of the rectangle will be rejected. To deal with the border conflicts the rectangles are shifted $L/3$ and $2L/3$ and the algorithm is run one more time for each shift. During these shifts, the algorithm explores regions where $|b| > 20 \Unit{deg}$, so clusters in that region might appear. Then the clusters found in the second or third run are only taken into account if none of its members is on any cluster of the previous runs; in this way we ensure that no clusters are missed or detected more than once because they are on the borders of the regions .

The last step in the preprocessing is the scaling of the star parameters used by DBSCAN. The algorithm makes use of the distance between sources in the N-dimensional space to define if the stars are clustered or not. Because there is no dimension preferred in the $5$-D parameter space ($l$, $b$, $\parallax$, $\mu_{\alpha^*}$, $\mu_{\delta}$), we standardize the parameters (rescale them to mean zero and variance one) so their weights in the process are equalized.

\subsection{DBSCAN}
\label{subsec:dbscan}

Once the region of the search is defined and the average distance between stars in the parameter space is determined, an automatic search for groups of stars that form an overdensity in the $5$-D space is started.

The clustering algorithm DBSCAN \citep{dbscan} is a density based algorithm that makes use of the notion of distance between two sources in the data to define a set of nearby points as a cluster; it has the advantage over other methods of being able to find arbitrary shaped clusters. An OC naturally falls in this description: they are groups of stars with common origin, so they share a common location $(l, b, \parallax)$ and motion $(\mu_{\alpha^*}, \mu_\delta)$. The TGAS \citep{gdr1-tgas} data set contains precise information for these five parameters, so one can define the distance between two stars \mbox{($i$ and $j$)} as 
\begin{align}
	&d(i,j) = \nonumber \\
	&\mbox{\small $\sqrt{(l_i-l_j)^2+(b_i-b_j)^2+(\parallax_i-\parallax_j)^2+(\mu_{\alpha^*,i}-\mu_{\alpha^*,j})^2+(\mu_{\delta,i}-\mu_{\delta,j})^2}$}.
\end{align} 
The choice of this euclidean distance is due to its simplicity, although a distance with specific weights on the different parameters in order to optimize the search for different kind of clusters (rich or poor, sparse or compact, etc) or to take into account the uncertainities of each value could be investigated. Notice also that the distance is calculated with the standardized values of these parameters. 

\begin{figure}[htb]
	\centering
	\includegraphics[width = 1\columnwidth]{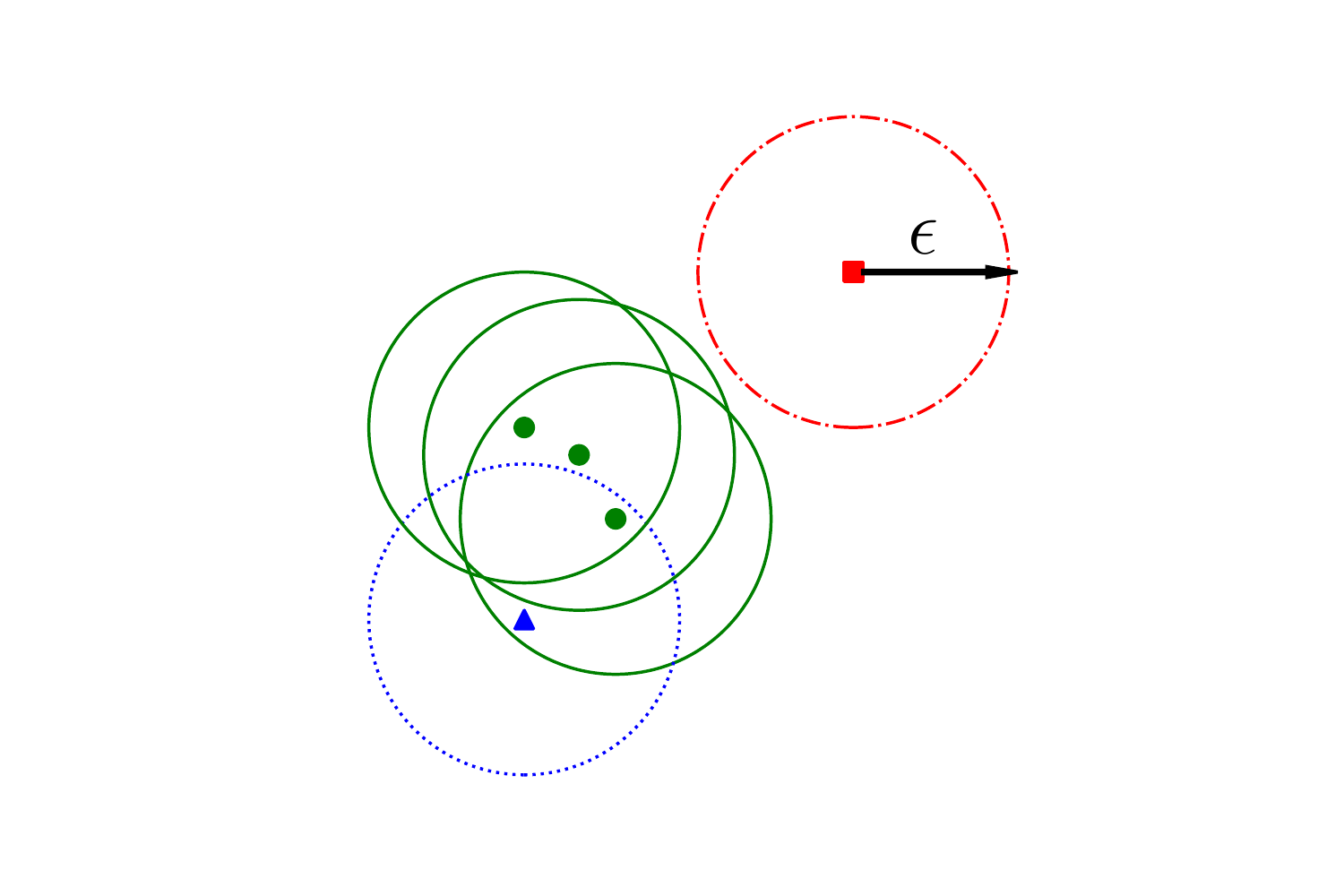}
	\caption{Schematic representation of a DBSCAN cluster with \textit{minPts}$ = 3$. Points in green represent core points, each point has \textit{minPts} points in its (green solid) hypersphere. The blue triangle represents a member point, it doesn't have \textit{minPts} in its (blue dashed) hypersphere but it is reached by a core point. The red square represents a field star, it does not have any other point in its (red dash-dot) hypersphere. All the hyperspheres have radius equal to $\epsilon$.}
	\label{fig:2d-dbscan}
\end{figure}

The definition of a DBSCAN cluster depends on two paramters: $\epsilon$ and \textit{minPts}. A hypersphere of radius $\epsilon$ is built centered on each source, and if the number of sources that fall inside the hypersphere is greater or equal than the pre-set \textit{minPts} the points are considered to be clustered. This definition of cluster allows to make the distinction between three types of sources in the data set: i) core points: sources that have a number of neighbours (within the hypersphere of radius $\epsilon$) greater or equal than \textit{minPts}, ii) members: sources that don't have these neighbours in their hyperspheres but they fall in the hypersphere of a core point, and iii) field stars: sources than do not fulfil any of the two previous conditions. For an intuitive $2$-Dimensional description of a cluster in DBSCAN see \figref{fig:2d-dbscan}.

\subsubsection*{Determination of the $\epsilon$ and \textit{minPts} parameters}
\label{subsubsec:determinationparameters}

Therefore the DBSCAN algorithm depends only on two parameters, the minimum number of sources (\textit{minPts}) to consider that a cluster exists and the radius ($\epsilon$) of the hypersphere where to search for these \textit{minPts} sources. In order to determine the optimum value of \textit{minPts} for OC detection the algorithm is tested with a simulated sample and a set of the values that perform best is chosen (see \secref{sec:simu}). In particular, the determination of $\epsilon$ is crucial for the efficiency of the detection, and the selected values can affect the number and shape of the clusters found.

\begin{figure}[htb]
	\centering
	\includegraphics[width = 1\columnwidth]{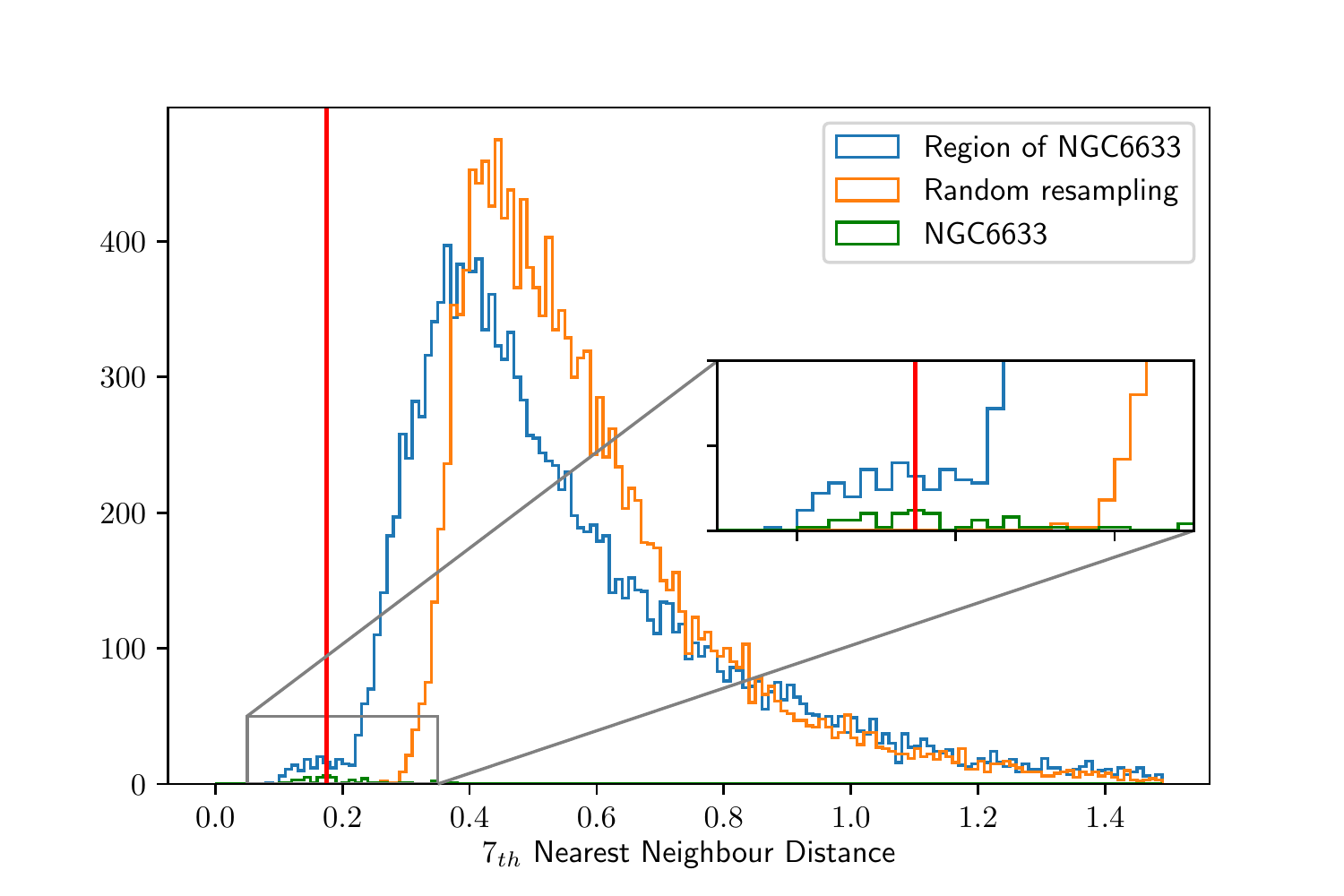}
	\caption{Histogram of the $7_{th}$-NNDs of the region around the cluster NGC6633. The blue line shows the $7_{th}$-NND histogram of all the stars in that sky region in TGAS. Orange line shows the $7_{th}$-NND histogram of one realization of a random resample. Green line shows the $7_{th}$-NND histogram for the listed members of NGC6633 (more visible in the zoom plot). The red line corresponds to the chosen value of $\epsilon$ in this region. The plot was made with the parameters $L = 14\Unit{deg}$ and \textit{minPts} $= 8$.}
	\label{fig:knn}
\end{figure}

Aiming to reduce the free input parameters we have implemented an automated determination of the $\epsilon$ value that best fits the data on a given region. Since a cluster is a concentration of stars in the parameter space, the distance of each star belonging to a cluster to its $k_{\rm th}$ nearest neighbour should be smaller than the average distance between stars belonging to the field (\figref{fig:knn}). Our determination of $\epsilon$, taking advantage of this fact, is as follows:

\begin{itemize}
	\item Compute the $k_{\rm th}$ Nearest Neighbour Distance ($k$NND) histogram for each region and store its minimum as $\epsilon_{\text{$k$NN}}$.
	\item Generate a new random sample, of the same number of stars, according to the distribution of each astrometric parameter estimated using a Gaussian kernel density estimator. Then, compute the $k$NND histogram for these stars and store the minimum value as $\epsilon_{\text{rand}}$. Since we are generating random samples, the minimum number of the $k$NND distribution will vary at each realization, in order to minimize this efect we store as $\epsilon_{\text{rand}}$ the average over $30$ repetitions of this step.
	\item Finally, to get the most concentrated stars and minimize the contamination from field stars, the choice of the parameter is $\epsilon = (\epsilon_{\text{$k$NN}} +\epsilon_{\text{rand}})/2$. 
\end{itemize}

 \figrefalt{fig:knn} shows a real distribution of $7_{th}$-NND around the cluster NGC6633 (in blue) together with a random resampled $7_{th}$-NND histogram (in orange) with the choice of $\epsilon$ in that region (red line), the peak belonging to the cluster is well separated from field stars through $\epsilon$. In addition it shows the histogram of distances to the $7_{th}$ neighbour of each star in the NGC6633 cluster (in green), where the members are taken from \cite{floor_clusters_DR1}.

The choice of the value for $k$ has to be related to the expected members of the cluster. Here, since \textit{minPts} determines the minimum members of a cluster, the value for $k$ is set to $k = \text{\textit{minPts}}-1$. Two free parameters ($L$, \textit{minPts}) are left to be optimized using simulations (see \secref{sec:simu}).

\subsection{Identification of Open Clusters}
\label{sec:identification}

At this point, when DBSCAN has found a list of candidate OCs, the method needs to be refined to distinguish real OCs from the statistical clusters (random accumulation of points). This step is an automatization of what is usually done by visual inspection; plot the Color-Magnitude Diagram (CMD) of the sky region and see if the clusterized stars follow an isochrone. We treat this as a pattern recognition problem, where Artificial Neural Networks (ANN) with a multilayer perceptron architecture have shown to be a good approach \citep{CMB-1995,RODPEHDGS-2004}. Similar problems, such as the identification of Globular Clusters \citep{MBSCMP-2012}, or a selection for QSO \citep{CYPPJR-2010} were also faced using a multilayer perceptron.

\subsubsection{Artificial Neural Networks}
\label{subsec:ann}

Artificial Neural Networks are computing models that try to mimic how a biological brain works. In particular, the multilayer perceptron consists in a set of at least three layers of nodes (neurons) capable to classify a given input feature vector into the class it belongs.  

\figrefalt{fig:ann} shows an schematic representation of a multilayer perceptron with one hidden layer. The left most (input) layer represents the set of input features $\{x_1,x_2,\ldots,x_n\}$. This is followed by the hidden layer, where each hidden neuron (labeled as $h_i$) weights the recieved input from the previous layer as $v_i = \omega_{i1} x_1 + \omega_{i2} x_2 + \ldots + \omega_{in} x_n$, and responds according to an activation function, in our case we use an hyperbolic tangent activation function 
\begin{equation}
y(v_i) = \text{tanh}(v_i),
\end{equation} 
which is then passed to the output layer that performs the classification.

\tikzstyle{cloud} = [draw,circle,fill=white]
\tikzstyle{line} = [draw, -latex']
\begin{figure}[htb]
\centering
\begin{tikzpicture}[node distance = 2cm, auto]
\node [cloud] at (0,0) (in1) {$x_1$};
\node [cloud] at (0,-1.5) (in2) {$x_2$};
\node [cloud] at (0,-4.5) (inn) {$x_n$};
\node at ($(in2)!.5!(inn)$) {\vdots};

\node [cloud] at (2,-0.75) (a1) {$h_1$};
\node [cloud] at (2,-1.75) (a2) {$h_2$};
\node [cloud] at (2,-3.75) (an) {$h_k$};
\node at ($(a2)!.5!(an)$) {\vdots};

\node [cloud] at (4,-2.25) (o1) {out};

\path [line] (in1) -- (a1);
\path [line] (in1) -- (a2);
\path [line] (in1) -- (an);
\path [line] (in2) -- (a1);
\path [line] (in2) -- (a2);
\path [line] (in2) -- (an);
\path [line] (inn) -- (a1);
\path [line] (inn) -- (a2);
\path [line] (inn) -- (an);
\path [line] (a1) -- (o1);
\path [line] (a2) -- (o1);
\path [line] (an) -- (o1);
\end{tikzpicture}
\caption{Schematic representation of a multilayer perceptron with one hidden layer. The $x_i$ values represent the input data. The $h_i$ labels represent neurons at the hidden layer.}
\label{fig:ann}
\end{figure}
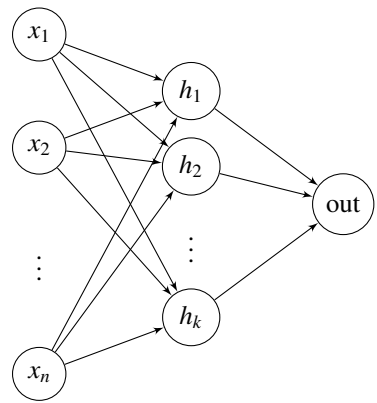
		
\subsubsection{Data Preparation}
\label{subsec:dataprep}

Artificial Neural Networks are supervised classification algorithms, they require a pre-classified learning sample for its training. In our case the data used to train the model are the Open Clusters taken from \cite{floor_clusters_DR1}. These clusters are well-characterized, having a reasonable number of members and showing clear isochrones in the CMD, that are the target of our pattern-recognition algorithm. Furthermore, they have the same astrometric uncertainties that our data so they are representative of our problem. In order to train the model, and to increase the size of the training set, several subsets of these OC member stars are randomly selected and plotted in a CMD to serve as patterns. Moreover, CMDs that do not correspond to clusters are needed too, as examples of negatives for the training. In this case we inspect the output from DBSCAN (for pairs of $(L,\text{\textit{minPts}})$ that were not used in the detection step) and select sets of clusterized stars not following any isochrone. 

\begin{figure}[htb]
\begin{subfigure}{1.00\columnwidth}
\centering
\includegraphics[scale = .5]{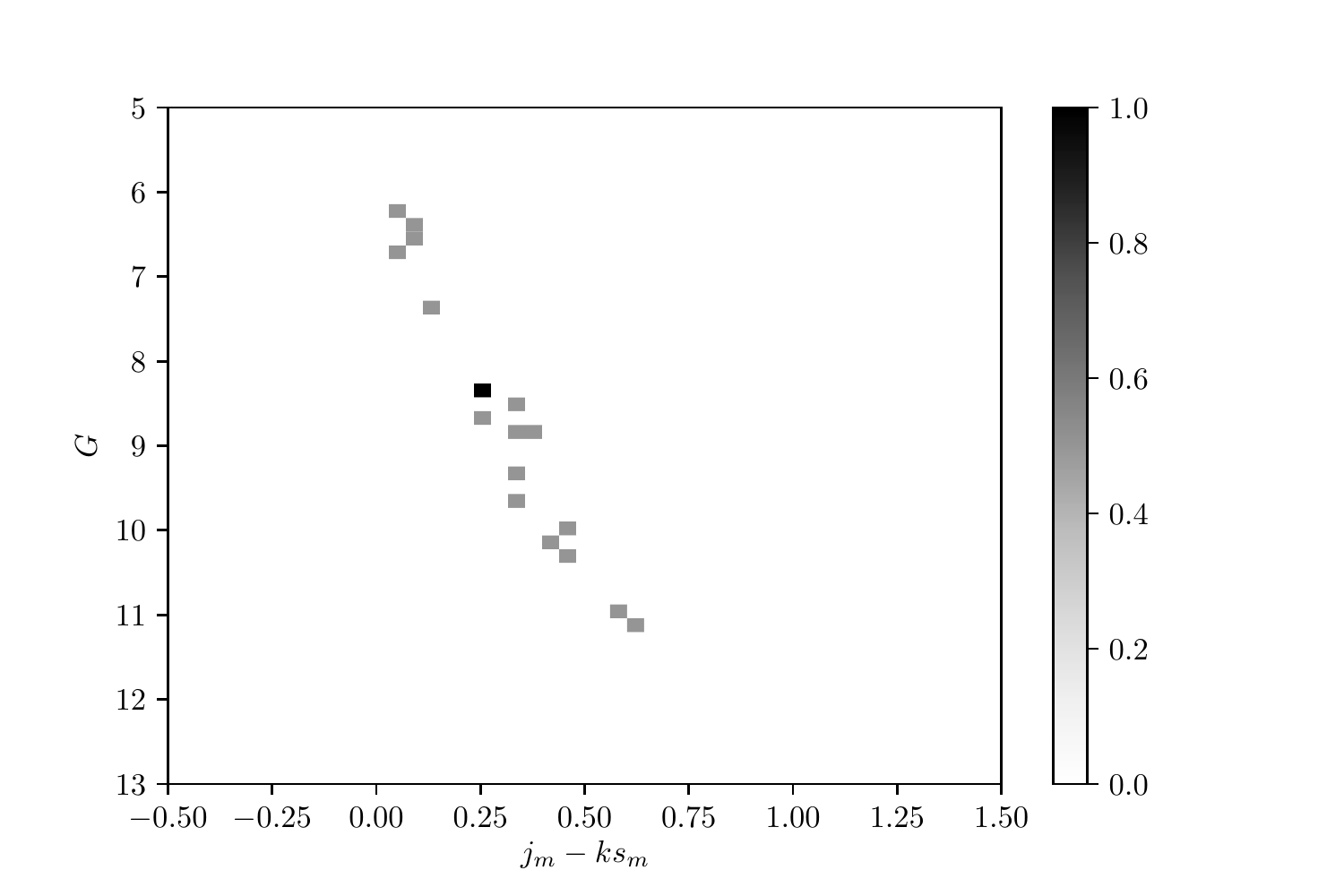}
\end{subfigure}
\begin{subfigure}{1.00\columnwidth}
\centering
\includegraphics[scale = .5]{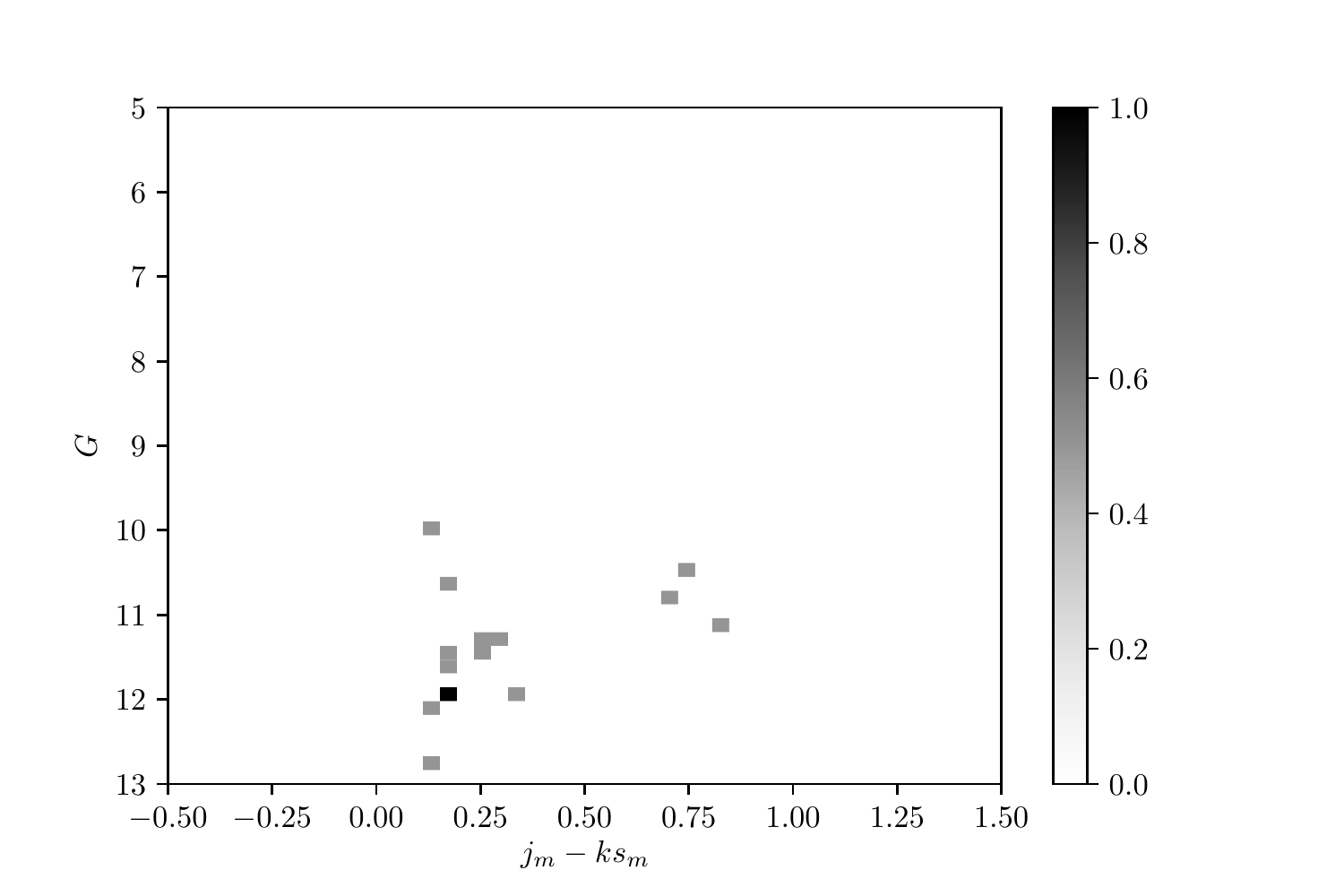}
\end{subfigure}
\caption{Examples of training data for the Artificial Neural Network classificator. The upper plot corresponds to a density map of a CMD of a subset of the members of Coma Berenices. The lower plot is the density map of a CMD of a cluster found by DBSCAN that we labeled as noise. In both cases the color represents the value of each pixel, and it is the input of the ANN model.}
\label{fig:CMDtraining}
\end{figure}

\figrefalt{fig:CMDtraining} shows two examples of training data for the model. The upper plot corresponds to members of the Coma Berenices cluster listed in \cite{floor_clusters_DR1}. The members are randomly chosen to form a set of $10$ sub-clusters, each one with characteristics similar to those found by DBSCAN. The CMD of these sub-clusters is then converted to a density map, so the value of each pixel can be used as the input of the ANN. A density map of one of these sub-clusters is shown in the upper plot. The lower plot corresponds to non-clusters for the training on negative identifications.

\subsubsection{Performance of the classification}
\label{subsec:performance}

The ANN classificator is trained with a total of $296$ images, containing a balanced relation between CMDs from true (real) OCs and CMDs from field stars. For performance estimation purposes this whole set is divided into a training and a test set, with $67\%$ and $33\%$ of the images each. The test CMDs are classified with a precision of a $97.95\%$ to the right class (OC or field stars). Even though the model is then trained with all the $296$ CMDs, the precision reached in the test set is only an estimation of the upper limit because the ANN  has learnt from the OCs in \gdrone listed in \cite{floor_clusters_DR1}. The detection of new OCs is then limited to have the same characteristics than those on \cite{floor_clusters_DR1}, where there are a total of $19$ nearby OCs with ages ranging from $40$ to $850$ Myr, and no significant differential extinction. A larger and wider, in terms of characteristics of the OCs, training set needs to be built in order to apply the method to the \gdrtwo data.

\section{Simulations}
\label{sec:simu}
	
A simulation of TGAS-like data is used to test the clustering method and set the optimal parameters to detect as many clusters as possible with a minimum of false positives.

The simulation consists, as described in \citet{dr1-validation}, in astrometric data from \tyctwo stars taken as nominal where errors coming from the AGIS solution have been added. The proper motions used for the simulation are those from \textit{Tycho}-2; to avoid their dispersion to spuriously increase when adding the TGAS errors they were "deconvolved" using Eq. 10 from \cite{FAXL-1999}. In the case of the parallaxes, for nearby stars the simulated value is a weighted average of "deconvolved" \hip parallaxes, while for the more distant stars it is taken from the photometric parallax in the \cite{APED-2011} catalogue. The simulation of the TGAS-like errors follows the description from \cite{tgas}, which is based on the algorithms from \cite{LLULDH-2012}. In short, this dataset is very representative of the real TGAS dataset that we will use both in terms of its distribution of parameters (taken from Tycho) and its astrometric errors (generated to be as close as possible to the TGAS ones).

The OCs are added to this dataset \textit{a posteriori}, simulated using the \gaia Object Generator \citep[GOG][]{gog} \citep[see details of how they are simulated in][]{OCsimulation}. For each cluster, the stars with $G>12$ are filtered out because the limiting magnitude in TGAS. Moreover, the simulation provides true values for the astrometric parameters to which observational errors are added. Using the uncertainties published in the TGAS catalogue, a normal random number is drawn centred in the true value, to compute the observed quantities.

\subsection*{Choice of the parameters}
\label{choiceparameters}

The selection to find the best parameters to run the algorithm is made in terms of noise and efficiency. Their definition, in terms of true positive rate (tp), false positive rate (fp) and false negative rate (fn), is fp/tp for noise and fn/tp for the efficiency.

\begin{figure*}[htb]
\begin{subfigure}{.33\textwidth}
	\centering
	\includegraphics[scale = 0.45]{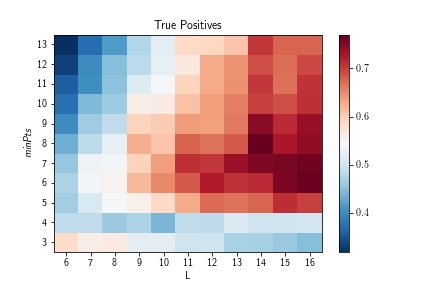}	
	\label{fig:sfig1}
\end{subfigure}%
\begin{subfigure}{.33\textwidth}
	\centering
	\includegraphics[scale = 0.45]{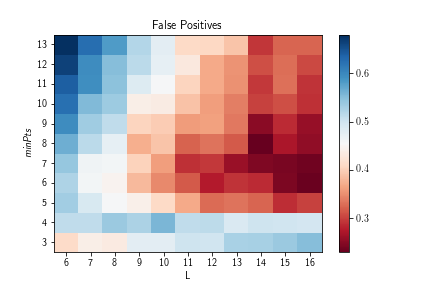}	
	\label{fig:sfig2}
\end{subfigure}
\begin{subfigure}{.34\textwidth}
	\centering
	\includegraphics[scale = 0.45]{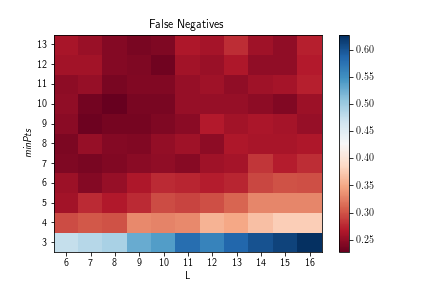}	
	\label{fig:sfig2}
\end{subfigure}
\begin{subfigure}{.5\textwidth}
	\centering
	\includegraphics[scale = 0.45]{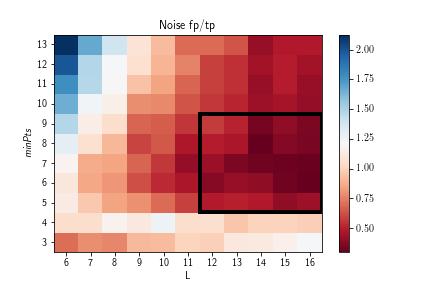}	
	\label{fig:sfig2}
\end{subfigure}
\begin{subfigure}{.5\textwidth}
	\centering
	\includegraphics[scale = 0.45]{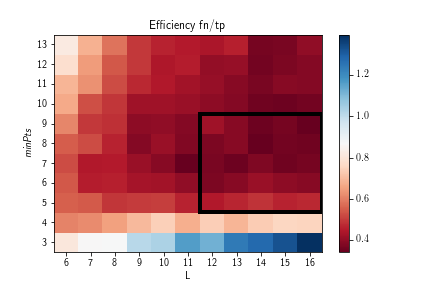}	
	\label{fig:sfig2}
\end{subfigure}
\caption{Performance of the algorithm with different set of parameters $(L, \text{\textit{minPts}})$ tested with simulated data. The top pannels represent the true positive (left), false positive (middle) and false negative (right) rates. Note the inversion of the color bar in the true positive rate to always represent the reddest pixels as the best perfoming pair of parameters. The bottom pannels represent noise (left) and efficiency (right). The black box encloses the area of pixels corresponding to the selected pairs of parameters.}
\label{fig:paramchoice}
\end{figure*}

In order to find the pairs of parameters that best perform, the algorithm was run over several pairs of $(L, \text{\textit{minPts}})$. The sweep over this parameter space allowed us to select the set of pairs of parameters that are less contaminated by spurious clusters. \figrefalt{fig:paramchoice} shows the performance of each pair of $(L, \text{\textit{minPts}})$, for the investigated pairs. The reddest pixel represents the best performing pairs of parameters while the bluest pixels represent the worst performing pairs. In the best case, with noise around $\sim0.25$, we are introducing $1$ spurious cluster in the detection every $4$ real clusters, while in the worst case we have a noise around $0.5$; and for efficiency $\sim0.3$, we don't detect one out of four real clusters. The selection was made trying to find a balance between noise and efficiency, the black box in \figref{fig:paramchoice} represents the selected pairs of $(L, \text{\textit{minPts}})$, which are $L \in [12,16]$ and $\text{\textit{minPts}} \in [5,9]$.

\section{Results}
\label{sec:results}

The whole method is run over the TGAS data to obtain a list of OC candidates. First, the DBSCAN algorithm is applied to the preselected data (see \secref{subsec:preprocessing}) with the optimal values for the parameters $L = \{12,13,14,15,16\}$ and \mbox{\textit{minPts} $ = \{5,6,7,8,9\}$}. This results in a list of clusterized stars, including real clusters already catalogued, non-catalogued possible clusters and noise. The clusters which are already catalogued, although they are useful to verify that the algorithm is capable of finding real clusters, are discarded (see \figref{fig:flowchart}). To do this, all the clusters found by DBSCAN whose centre lies within a box of $2\Unit{deg}\times2\Unit{deg}$ centred in a cluster present in the MWSC catalogue are discarded. In this way we ensure a list composed only of new cluster candidates.
\cite{SRESBG-2016} published a list of nine nearby open clusters using proper
motions from a combination of Tycho-2 with URAT1 catalogues. We did not include these
clusters in the cross-match with known clusters step in order to use them as a check of the method.

The classification of these clusters into probable OC candidates and statistical clusters is done with the ANN algorithm. The model is trained with CMDs from real clusters (see \secref{subsec:dataprep}) with the phometric data from $2$MASS and TGAS, and it is capable to identify isochrone patterns in CMDs. The isochrone patterns identified by the ANN model are based on those of the OCs listed in \cite{floor_clusters_DR1}. Only the clusters found to follow an isochrone with a confidence level higher than $90\%$ are selected.

\subsection{Open Cluster Candidates}
\label{subsec:OCcand}

\tabref{tab:oclist} lists $31$ Open Cluster candidates resulting from the application of the above described algorithms. We include the mean sky position, proper motions and parallaxes of the identified members. We do not provide uncertainties because the data has been superseeded by \gaia DR2. Because the method is run over $25$ different pairs of parameters $(L,\text{\textit{minPts}})$, the final list is sorted by the number of appearances of the clusters in the different pairs of parameters. The value $N_{\rm found}$ indicates how many times the cluster has been found for the used pairs of $(L,\text{\textit{minPts}})$.

As mentioned above, we did not include the OCs in \cite{SRESBG-2016} in the list of previously known clusters and therefore we expect some overlap with our candidates. This is the case of our UBC1 and UBC12, which are RSG4 and RSG3, respectively.

Of the other seven clusters, RSG2 was not found possibly because its high galactic latitude and its high $\mu_{\delta}$ mean, which is $-29.54\Unit{mas\cdot yr}^{-1}$. Because our preprocessing removes stars with $|\mu_{\alpha^*}|,|\mu_{\delta}| > 30\Unit{mas\cdot yr}^{-1}$ (see \secref{subsec:preprocessing}) and due to the proper motion uncertainties in the TGAS catalogue, we may lose part of the members and, so, the algorithm does not consider the surviving members as a cluster. On the other hand, the criterium to match our candidates
with the list of known OCs is purely positional (within a box of 2 deg $\times$ 2 deg).
We do not impose a match in proper motions and/or parallaxes because the large
differences between the values quoted in MWSC and \cite{dias}, which makes us doubt about
the reliability of some values. This criterion discards candidate clusters that are
in the vicinity of know clusters. And, this is the case of RSG1, RSG5, RSG6, RSG7, RSG8 and RSG9.
Our candidate list is therefore not complete, specially at very low latitudes where the density of
known clusters increases.

UBC7 shares proper motions and parallax with Collinder~135. It is located at 2.3 deg of the quoted Collinder~135 center and for this reason it is not matched in our step to discard already known clusters. Figure 7 shows a cone search of 10 deg centred in UBC7 where a pattern in the data is clearly visible. This pattern is an artefact of the Gaia scanning law in the 14 months mission of Gaia DR1. UBC7 is located where two stripes cross and this, together with the fact that their stars also share parallaxes and proper motions, leads to its detection as a separate cluster. This is an indication that the inhomogeneities in the sky coverage of TGAS data might lead to the detection of spurious clusters. Collinder~135 is not detected by DBSCAN because their members lie in a region not well covered by the observations, they are more spread than UBC7 and they are not recognized as a group. It can happen that Collinder~135 is larger than quoted in the literature and includes UBC7.

\begin{figure}[htb]
\centering
\includegraphics[width = 1.\columnwidth]{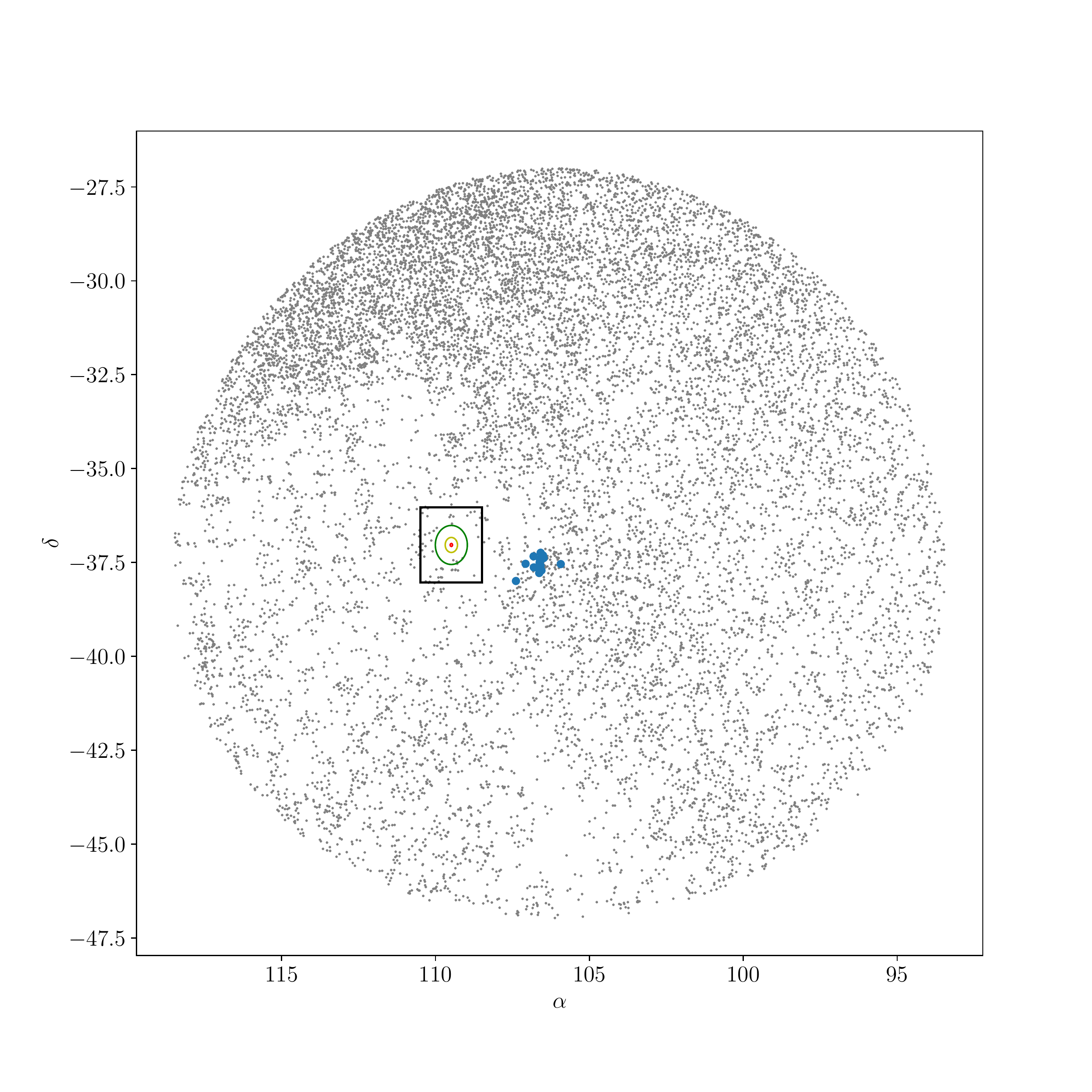}
\caption{Cone search of $10 \Unit{deg}$ centred in UBC7 in the TGAS data with
	 more than 120 photometric observations. Blue dots represent members of UBC7. 
	 The red, yellow and green circles represent the $r_0$, $r_1$ and $r_2$ radius in the MWSC catalogue 
	 for Collinder 135. The black box is the 2 deg $\times$ 2 deg zone where all candidate clusters are
	 considered as known clusters. The visible stripes on the data are due to the \gaia scanning law.}
\label{fig:collinder135}
\end{figure}

\begin{table*}
\caption{List of the 31 Open Cluster candidates. The parameters are the mean of the members found with TGAS. $N_{\rm found}$ refers to the times each cluster has been found within the explored parameters $(L,\text{\textit{minPts}})$. UBC stands for University of Barcelona Cluster.}
\label{tab:oclist}
\centering
\begin{threeparttable}[b]
\begin{tabular}{ccccccccc}
\hline
\hline
\multicolumn{1}{c}{Name} &  \multicolumn{1}{c}{\begin{tabular}[c]{@{}c@{}}$\alpha$\\ $[\Unit{deg}]$\end{tabular}} &\multicolumn{1}{c}{\begin{tabular}[c]{@{}c@{}}$\delta$\\ $[\Unit{deg}]$\end{tabular}} & \multicolumn{1}{c}{\begin{tabular}[c]{@{}c@{}}$l$\\ $[\Unit{deg}]$\end{tabular}} & \multicolumn{1}{c}{\begin{tabular}[c]{@{}c@{}}$b$\\ $[\Unit{deg}]$\end{tabular}} & \multicolumn{1}{c}{\begin{tabular}[c]{@{}c@{}}$\parallax$\\ $[\Unit{mas}]$\end{tabular}} & \multicolumn{1}{c}{\begin{tabular}[c]{@{}c@{}}$\mu_{\alpha^*}$\\ $[\Unit{mas\cdot yr}^{-1}]$\end{tabular}} & \multicolumn{1}{c}{\begin{tabular}[c]{@{}c@{}}$\mu_{\delta}$\\ $[\Unit{mas\cdot yr}^{-1}]$\end{tabular}}  & \multicolumn{1}{c}{$N_{\rm found}$} \\
\hline
UBC1\tnote{a}	&	$287.83$	&	$56.62$		&	$87.30$		&	$19.77$		&	$3.04$	&	$-2.80$	&	$3.69$	&	$27$	\\
UBC2	&	$4.90$		&	$46.38$		&	$117.22$	&	$-16.13$	&	$1.62$	&	$-5.95$	&	$-5.67$	&	$24$	\\
UBC3	&	$283.74$	&	$12.29$		&	$44.29$		&	$4.80$		&	$0.53$	&	$-1.57$	&	$-2.31$	&	$21$	\\
UBC4	&	$60.73$		&	$35.23$		&	$161.37$	&	$-12.97$	&	$1.74$	&	$-0.08$	&	$-5.36$	&	$21$	\\
UBC5	&	$238.65$	&	$-47.66$	&	$331.90$	&	$4.63$		&	$1.61$	&	$-7.21$	&	$-4.80$	&	$18$	\\
UBC6	&	$343.87$	&	$51.14$		&	$105.06$	&	$-7.65$		&	$1.35$	&	$-7.46$	&	$-4.54$	&	$15$	\\
UBC7\tnote{b}	&	$106.64$	&	$-37.54$	&	$248.52$	&	$-13.36$	&	$3.67$	&	$-9.43$	&	$7.03$	&	$14$	\\
UBC8	&	$84.65$		&	$56.99$		&	$155.06$	&	$13.35$		&	$2.17$	&	$-3.35$	&	$-3.24$	&	$13$	\\
UBC9	&	$276.60$	&	$26.42$		&	$54.48$		&	$16.84$		&	$2.80$	&	$-0.12$	&	$-5.31$	&	$12$	\\
UBC10	&	$324.20$	&	$60.86$		&	$101.34$	&	$6.43$		&	$0.99$	&	$-1.73$	&	$-3.15$	&	$10$	\\
UBC11	&	$246.61$	&	$-60.17$	&	$326.80$	&	$-7.69$		&	$2.15$	&	$-0.25$	&	$-7.34$	&	$10$	\\
UBC12\tnote{c}	&	$126.11$	&	$-8.39$		&	$231.65$	&	$16.32$		&	$2.32$	&	$-8.19$	&	$4.47$	&	$6$	\\
UBC13	&	$121.24$	&	$4.14$		&	$217.71$	&	$18.23$		&	$1.75$	&	$-7.22$	&	$-1.48$	&	$5$	\\
UBC14	&	$295.01$	&	$3.21$		&	$41.43$		&	$-9.29$		&	$1.33$	&	$0.56$	&	$-1.76$	&	$5$	\\
UBC15	&	$268.05$	&	$-25.89$	&	$3.35$		&	$0.30$		&	$0.77$	&	$1.06$	&	$-1.38$	&	$4$	\\
UBC16	&	$143.77$	&	$-27.40$	&  	$258.09$ 	&  	$17.91$ 	& 	$1.93$ 	& 	$-4.67$ &  	$2.15$	&	$3$	\\
UBC17	&	$83.15$		&	$-1.57$		&	$205.11$ 	&  	$-18.20$ 	&  	$2.70$ 	& 	$-0.02$ & 	$-0.41$	&	$3$	\\
UBC18	&	$97.59$		&	$-39.65$	&	$247.88$ 	&   $-20.72$	&  	$1.40$ 	& 	$0.91$ 	& 	$6.70$	&	$2$	\\
UBC19	&	$56.63$		&	$29.93$		&	$162.35$ 	&  	$-19.22$ 	&  	$2.70$ 	&  	$2.39$ 	& 	$-4.56$	&	$2$	\\
UBC20	&	$278.66$	&	$-13.77$	&	$18.77$		&	$-2.59$		&	$0.50$	&	$-0.13$	&	$-2.13$	&	$2$	\\
UBC21	&	$130.06$	&	$-21.06$	&	$244.72$	&	$12.45$		&	$1.18$	&	$-6.13$	&	$2.40$	&	$2$	\\
UBC22	&	$90.00$		&	$14.14$		&	$194.46$	&	$-4.62$		&	$0.66$	&	$0.06$	&	$-2.93$	&	$1$	\\
UBC23	&	$252.57$	&	$-4.79$		&	$13.50$		&	$24.14$		&	$1.76$	&	$-4.41$	&	$-6.76$	&	$1$	\\
UBC24	&	$256.48$	&	$1.26$		&	$21.39$		&	$23.91$		&	$2.02$	&	$-3.66$	&	$-1.65$	&	$1$	\\
UBC25	&	$257.20$	&	$-17.50$	&	$4.98$		&	$13.31$		&	$1.20$	&	$-4.20$	&	$-4.87$	&	$1$	\\
UBC26	&	$285.49$	&	$22.05$		&	$53.83$		&	$7.66$		&	$1.63$	&	$2.07$	&	$-5.44$	&	$1$	\\
UBC27	&	$294.30$	&	$15.57$		&  	$51.98$ 	&  	$-2.72$ 	& 	$0.85$ 	& 	$-1.36$ &  	$-5.90$	&	$1$	\\
UBC28	&	$332.41$	&	$66.51$		&	$107.78$ 	&  	$8.53$ 		&  	$1.02$ 	& 	$-4.34$ & 	$-3.39$	&	$1$	\\
UBC29	&	$129.43$	&	$-16.54$	&	$240.57$	&  	$14.58$ 	&  	$1.21$ 	& 	$-6.38$ & 	$2.13$	&	$1$	\\
UBC30	&	$3.15$		&	$73.14$		&	$120.08$ 	&   $10.49$		&  	$1.12$ 	& 	$2.10$ 	& 	$0.62$	&	$1$	\\
UBC31	&	$61.06$		&	$32.14$		&	$163.74$ 	&  	$-15.04$ 	&  	$2.85$ 	&  	$3.69$ 	& 	$-5.04$	&	$1$	\\
\hline
\end{tabular}
\begin{tablenotes}
\item[a] is RSG4 in \citet{SRESBG-2016}
\item[b] probably related to Collinder 135
\item[c] is RSG3 in \citet{SRESBG-2016}
\end{tablenotes}
\end{threeparttable}
\end{table*} 

\section{Validation using \gdrtwo}
\label{sec:dr2val}

\gdrtwo provides an excellent set of data for the confirmation of our candidate members
because of the improved precision of the astrometric parameters, the availability of
those parameters for the stars down to $\sim21$ mag and the availability of precise
$G$, $G_{BP}$ and $G_{RP}$ photometry.

\begin{figure}[htb]
\centering
\includegraphics[width = 1.\columnwidth]{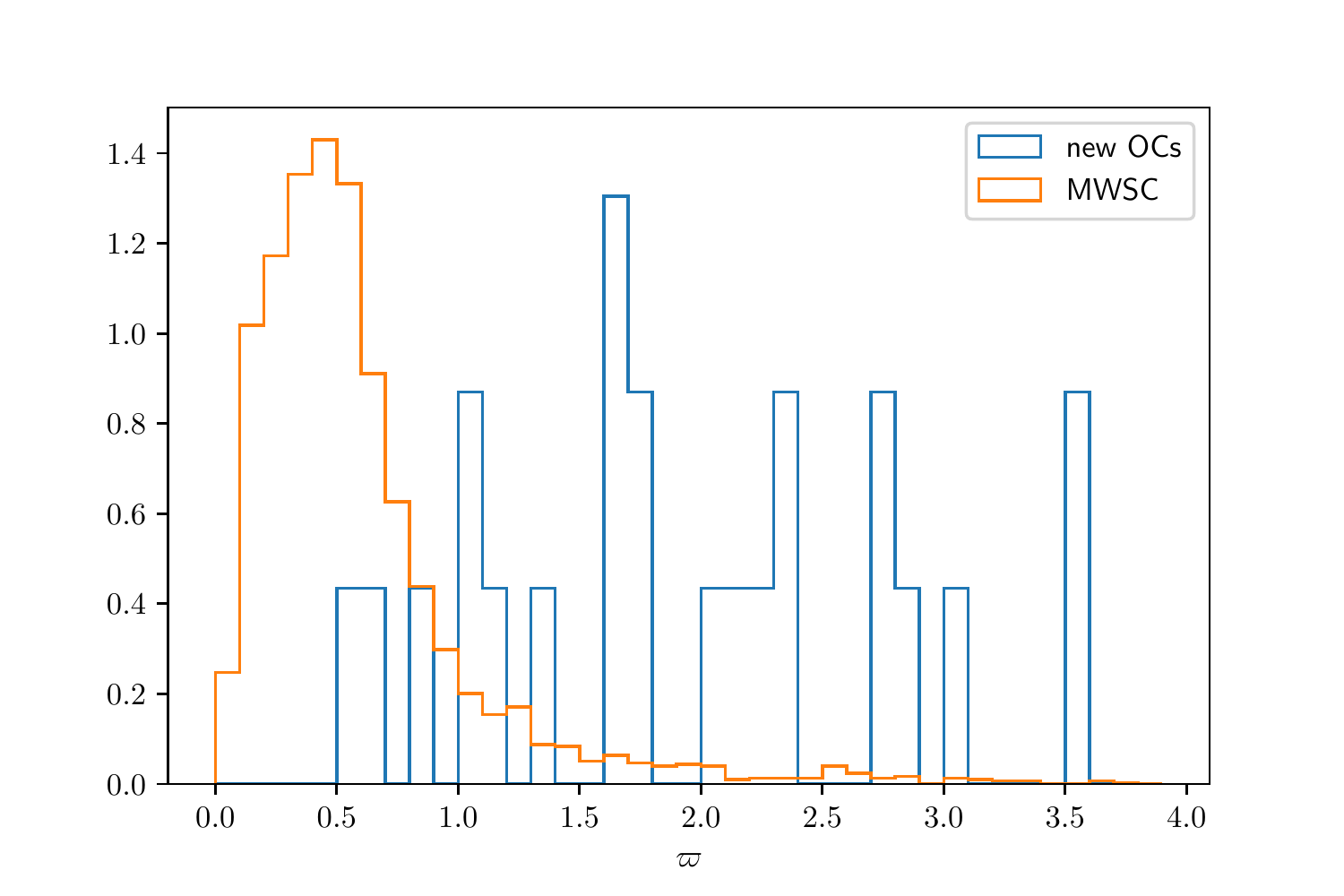}
\caption{Normalized parallax distribution of the found OCs (blue) and the ones listed in MWSC (orange). The newly detected OCs are closer than most of the catalogued clusters in MWSC.}
\label{fig:par_dist}
\end{figure}

In order to validate each cluster, we run again our method in a set of DR2 objects selected in a region around its centre (a cone search of $1$ or $2\Unit{deg}$ depending on the mean parallax of the cluster). The determination of the $\epsilon$ parameter for DBSCAN is now more complicated due to the higher density of stars in the \gdrtwo data, reaching in some studied cases $\sim \numprint{150000}$ stars in that region. Because our goal here is just to validate the already found candidates (not detecting new OCs) and thus validate our method, we apply a set of cuts in the data. These cuts are mainly in magnitude and parallax to increase the contrast between the cluster and field populations, to avoid large uncertainties, and to discard distant stars (being our candidates detected with TGAS data, the clusters are necessarily nearby, see \figref{fig:par_dist}).

\begin{figure*}[htb]
\centering
\includegraphics[width = 1.\textwidth]{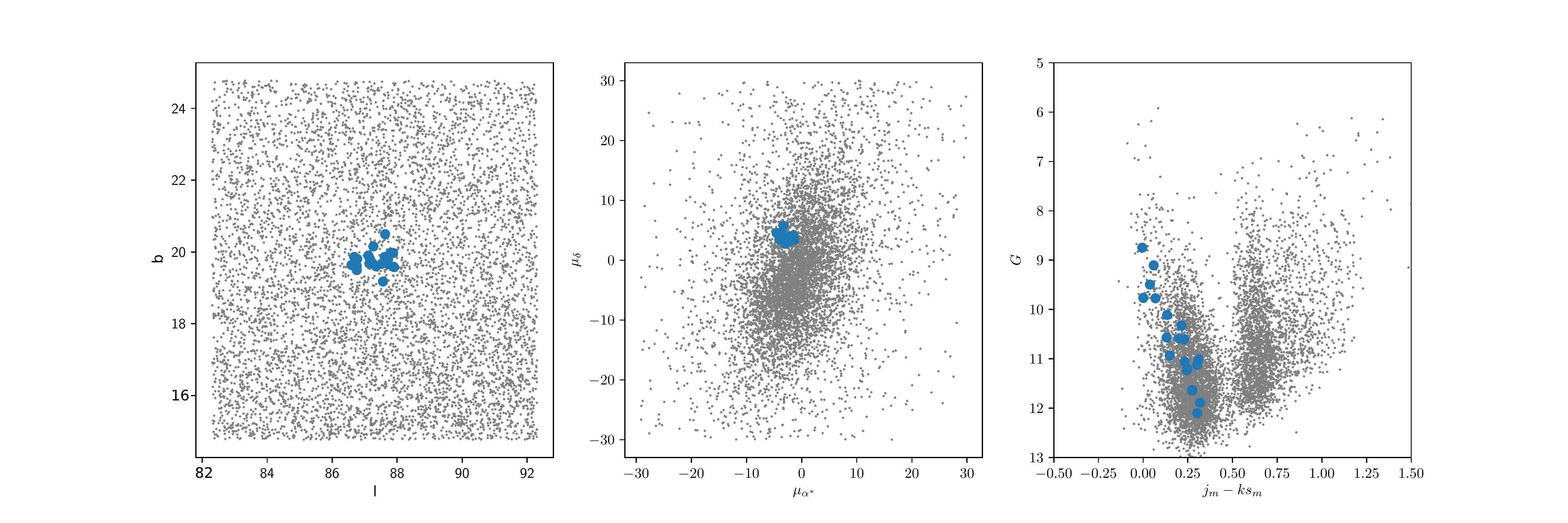}
\includegraphics[width = 1.\textwidth]{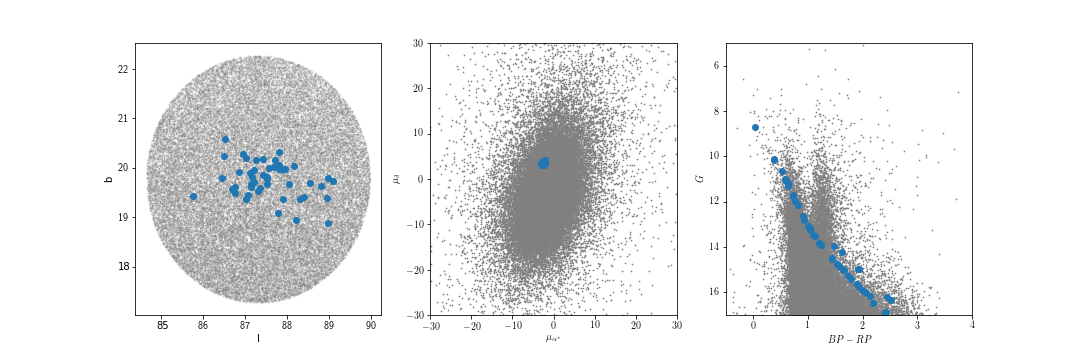}
\caption{Visualization of UBC1 from \tabref{tab:oclist}. In the top pannels, the left plot shows the position of the member stars (blue) along with field stars (grey) in a $10\Unit{deg}\times10\Unit{deg}$ area in TGAS data. Middle plot shows the same stars in the proper motion space. Right plot represents the CMD of the stars in the field using photometry from \gaia and $2$MASS, member stars follow an isochrone. The bottom pannels are the equivalent for \gdrtwo data. The major difference is in the CMD, where the members detected in \gdrtwo are clearly following an isochrone due to the better quality of the photometric \gaia data.}
\label{fig:cluster_summary_0}
\end{figure*}

\figrefalt{fig:cluster_summary_0} shows an example of UBC1 in the TGAS (top panels) and \gdrtwo (bottom panels) data. Left plots show the spatial distribution of the member stars found in each data set, in the TGAS case it shows a squared area of $10\Unit{deg} \times 10\Unit{deg}$ whilst in \gdrtwo it is a cone search of $2\Unit{deg}$. The middle plots show the members in the proper motion space and we can see that in \gdrtwo data the stars are more compact. The major difference is in the rightmost plots where a CMD is shown for both cases, one using photometry from $2$MASS (top) and one using only \gaia data (bottom). The much better quality of the \gaia photometric data (both plots share the same stars for $G \leq 12$) allow us to see much clearly the isochrone pattern that the member stars follow. 

We are able to re-detect, and thus confirm, a high percentage of the listed OCs using DBSCAN in a region around the cluster. \tabref{tab:oclist_confirmed} lists the confirmed OCs. The clusters that we consider as confirmed are those who share most of the stars with those previously found in TGAS. See plots similar to \figref{fig:cluster_summary_0} in Appendix~\ref{sec:app_cmd} for all the OCs. {\em Gaia} DR2 includes mean radial velocities for stars brighter than 12 mag. In \tabref{tab:oclist_confirmed} we include the mean radial velocity for the OCs derived from the identified members. 

\begin{table*}
\caption{List of the confirmed Open Clusters. The parameters are the mean (and standard deviation) of the members found with \gaia DR2. We also include radial velocity for those stars available. $N$ refers to the number of members found (and members to compute mean radial velocity).}
\label{tab:oclist_confirmed}
\centering
\resizebox{\textwidth}{!}{\begin{tabular}{cccccccccc}
\hline
\hline
\multicolumn{1}{c}{Name} &  \multicolumn{1}{c}{\begin{tabular}[c]{@{}c@{}}$\alpha$\\ $[\Unit{deg}]$\end{tabular}} &\multicolumn{1}{c}{\begin{tabular}[c]{@{}c@{}}$\delta$\\ $[\Unit{deg}]$\end{tabular}} & \multicolumn{1}{c}{\begin{tabular}[c]{@{}c@{}}$l$\\ $[\Unit{deg}]$\end{tabular}} & \multicolumn{1}{c}{\begin{tabular}[c]{@{}c@{}}$b$\\ $[\Unit{deg}]$\end{tabular}} & \multicolumn{1}{c}{\begin{tabular}[c]{@{}c@{}}$\parallax$\\ $[\Unit{mas}]$\end{tabular}} & \multicolumn{1}{c}{\begin{tabular}[c]{@{}c@{}}$\mu_{\alpha^*}$\\ $[\Unit{mas\cdot yr}^{-1}]$\end{tabular}} & \multicolumn{1}{c}{\begin{tabular}[c]{@{}c@{}}$\mu_{\delta}$\\ $[\Unit{mas\cdot yr}^{-1}]$\end{tabular}}  & \multicolumn{1}{c}{\begin{tabular}[c]{@{}c@{}}$V_{\rm rad}$\\ $[\Unit{km\cdot s}^{-1}]$\end{tabular}}  & \multicolumn{1}{c}{$N$ ($N_{V_{\rm rad}}$)} \\
\hline
UBC1	&	$288.00$ $(0.84)$	&	$56.83$ $(0.63)$	&	$87.55$ $(0.74)$	&	$19.76$ $(0.35)$	&	$3.05$ $(0.02)$	&	$-2.49$ $(0.25)$	&	$3.69$ $(0.24)$	&	$-21.46$ $(2.36)$	&	$47$ $(14)$	\\
UBC2	&	$5.80$ $(0.84)$		&	$46.59$ $(0.34)$	&	$117.89$ $(0.62)$	&	$-15.99$ $(0.32)$	&	$1.74$ $(0.03)$	&	$-6.34$ $(0.12)$	&	$-5.03$ $(0.13)$&	$-9.73$ $(2.22)$	&	$23$ $(4)$	\\
UBC3	&	$283.77$ $(0.16)$	&	$12.34$ $(0.22)$	&	$44.35$ $(0.24)$	&	$4.79$ $(0.12)$		&	$0.58$ $(0.04)$	&	$-0.60$ $(0.08)$	&	$-1.36$ $(0.09)$&	$-7.25$ $(13.54)$	&	$29$ $(2)$	\\
UBC4	&	$60.96$ $(1.07)$	&	$35.35$ $(0.74)$	&	$161.42$ $(1.05)$	&	$-12.75$ $(0.50)$	&	$1.64$ $(0.05)$	&	$-0.75$ $(0.13)$	&	$-5.72$ $(0.13)$&	$3.67$ $(1.65)$	&	$44$ $(3)$	\\
UBC5	&	$238.42$ $(0.74)$	&	$-47.72$ $(0.41)$	&	$331.74$ $(0.56)$	&	$4.68$ $(0.32)$		&	$1.78$ $(0.01)$	&	$-6.69$ $(0.15)$	&	$-4.18$ $(0.09)$&	$-14.91$ $(-)$		&	$29$ $(1)$	\\
UBC6	&	$343.95$ $(0.48)$	&	$51.19$ $(0.19)$	&	$105.13$ $(0.29)$	&	$-7.63$ $(0.21)$	&	$0.67$ $(0.01)$	&	$-4.64$ $(0.06)$	&	$-4.90$ $(0.08)$&	$-31.64$ $(1.51)$			&	$76$ $(3)$	\\
UBC7	&	$106.92$ $(0.61)$	&	$-37.74$ $(0.65)$	&	$248.80$ $(0.71)$	&	$-13.25$ $(0.42)$	&	$3.56$ $(0.05)$	&	$-9.74$ $(0.19)$	&	$6.99$ $(0.20)$	&	$16.42$ $(4.71)$			&	$77$ $(21)$	\\
UBC8	&	$84.36$	 $(0.86)$	&	$57.16$ $(0.54)$	&	$154.83$ $(0.64)$	&	$13.30$ $(0.36)$	&	$2.05$ $(0.03)$	&	$-3.14$	$(0.17)$	&	$-3.99$ $(0.16)$&	$-5.96$ $(3.94)$			&	$103$ $(21)$	\\
UBC9	&	$276.64$ $(0.41)$	&	$26.40$ $(0.39)$	&	$54.48$ $(0.40)$	&	$16.80$ $(0.38)$	&	$2.87$ $(0.02)$	&	$0.60$ $(0.16)$		&	$-5.35$ $(0.18)$&	$-17.98$ $(3.12)$			&	$25$ $(6)$	\\
UBC10a	&	$324.46$ $(1.36)$	&	$61.75$ $(0.95)$	&	$102.03$ $(1.02)$	&	$7.02$ $(0.55)$	&	$1.07$ $(0.01)$	&	$-2.14$ $(0.11)$		&	$-3.03$ $(0.12)$&	$-23.12$ $(-)$			&	$43$ $(1)$	\\
UBC10b	&	$326.87$ $(0.96)$	&	$61.10$ $(0.47)$	&	$102.49$ $(0.36)$	&	$5.75$ $(0.55)$	&	$1.01$ $(0.01)$	&	$-3.46$ $(0.09)$		&	$-1.86$ $(0.10)$&	$-46.90$ $(-)$			&	$40$ $(1)$	\\
UBC11	&	$246.16$ $(1.91)$	&	$-59.94$ $(0.87)$	&	$326.81$ $(1.15)$	&	$-7.39$ $(0.61)$	&	$2.13$ $(0.04)$	&	$-0.30$ $(0.37)$	&	$-6.78$ $(0.28)$&	$-18.18$ $(5.35)$			&	$44$ $(4)$	\\
UBC12	&	$126.13$ $(0.65)$	&	$-8.56$ $(0.47)$	&	$231.81$ $(0.71)$	&	$16.24$ $(0.41)$	&	$2.21$ $(0.05)$	&	$-8.27$ $(0.20)$	&	$4.07$ $(0.28)$	&	$31.34$ $(-)$			&	$19$ $(1)$	\\
UBC13	&	$120.90$ $(0.79)$	&	$3.60$ $(1.14)$		&	$218.04$ $(1.02)$	&	$17.68$ $(0.99)$	&	$1.60$ $(0.04)$	&	$-7.76$ $(0.19)$	&	$-1.16$ $(0.21)$&	$22.91$ $(5.48)$			&	$36$ $(6)$	\\
UBC14	&	$294.80$ $(0.58)$	&	$3.64$ $(1.01)$		&	$41.70$ $(1.06)$	&	$-8.91$ $(0.52)$	&	$1.30$ $(0.02)$	&	$0.14$ $(0.16)$		&	$-2.09$ $(0.20)$&	$-9.85$ $(-)$			&	$46$ $(1)$	\\
UBC17a	&	$83.38$	$(0.22)$	&	$-1.58$ $(0.86)$	&	$205.23$ $(1.04)$ 	&  	$-18.01$ $(1.06)$ 	&  	$2.74$ $(0.04)$ & 	$1.59$ $(0.27)$ 	& 	$-1.20$ $(0.35)$&	$18.96$ $(7.64)$			&	$180$ $(18)$	\\
UBC17b	&	$83.35$	$(0.76)$	&	$-1.54$ $(0.94)$	&	$205.18$ $(0.95)$ 	&  	$-18.02$ $(0.79)$ 	&  	$2.36$ $(0.04)$ & 	$0.05$ $(0.17)$ 	& 	$-0.16$ $(0.24)$&	$33.19$ $(4.41)$			&	$103$ $(4)$	\\
UBC19	&	$56.48$ $(0.37)$	&	$29.91$ $(0.22)$	&	$162.25$ $(0.24)$ 	&  	$-19.32$ $(0.32)$ 	&  	$2.39$ $(0.11)$ &  	$2.71$ $(0.53)$ 	& 	$-5.19$ $(0.27)$&	$31.38$ $(3.46)$			&	$34$ $(2)$	\\
UBC21	&	$130.35$ $(0.81)$	&	$-20.68$ $(0.94)$	&	$244.56$ $(1.10)$	&	$12.87$ $(0.55)$	&	$1.12$ $(0.02)$	&	$-6.51$ $(0.22)$	&	$2.48$ $(0.17)$&	$-$ $(-)$			&	$47$ $(0)$	\\
UBC26	&	$285.24$ $(0.69)$	&	$21.92$ $(0.74)$	&	$53.61$ $(0.86)$	&	$7.80$ $(0.49)$		&	$1.66$ $(0.03)$	&	$2.01$ $(0.17)$		&	$-5.18$ $(0.21)$&	$6.79$ $(17.43)$			&	$64$ $(2)$	\\
UBC27	&	$294.31$ $(0.25)$	&	$15.58$ $(0.25)$	&  	$52.00$ $(0.24)$ 	&  	$-2.73$ $(0.25)$ 	& 	$0.88$ $(0.03)$ & 	$-0.82$ $(0.07)$ 	&  	$-6.22$ $(0.08)$&	$-$ $(-)$			&	$65$ $(0)$	\\
UBC31	&	$61.11$ $(1.21)$	&	$32.76$ $(1.13)$	&	$163.33$ $(1.04)$ 	&  	$-14.55$ $(1.14)$ 	&  	$2.70$ $(0.07)$ &  	$3.77$ $(0.22)$ 	& 	$-5.43$ $(0.24)$&	$22.74$ $(5.73)$			&	$84$ $(12)$	\\
UBC32	&	$279.43$ $(0.66)$	&	$-14.04$ $(0.93)$	&	$18.87$ $(0.96)$	&	$-3.38$ $(0.60)$	&	$3.56$ $(0.04)$	&	$-1.75$ $(0.26)$	&	$-9.26$ $(0.29)$&	$-21.58$ $(7.24)$			&	$60$ $(14)$	\\
\hline
\end{tabular}}
\end{table*} 

The non-confirmed clusters are UBC15, UBC16, UBC18, UBC22, UBC23, UBC24, UBC25, UBC28, UBC29 and UBC30. They are all in the second half of \tabref{tab:oclist}, which means the less frequently found ($N_{\rm found} < 5$) within the explored parameters $(L,\text{\textit{minPts}})$. The criteria followed in order to sort the list of candidates is reasonable; a $100\%$ of the clusters with $N_{\rm found} \geq 5$ are confirmed, while for $N_{\rm found} < 5$ has a $59\%$ of confirmation. As a whole, we are able to confirm $\sim 70\%$ of the proposed candidates; this is within the expected performance limits obtained in the simulations, where we have around $25\%$ and $50\%$ in terms of noise (see \secref{choiceparameters}).

\subsection {Comments on some of the confirmed clusters}

\begin{figure*}[htb]
\centering
\includegraphics[width = 1.\textwidth]{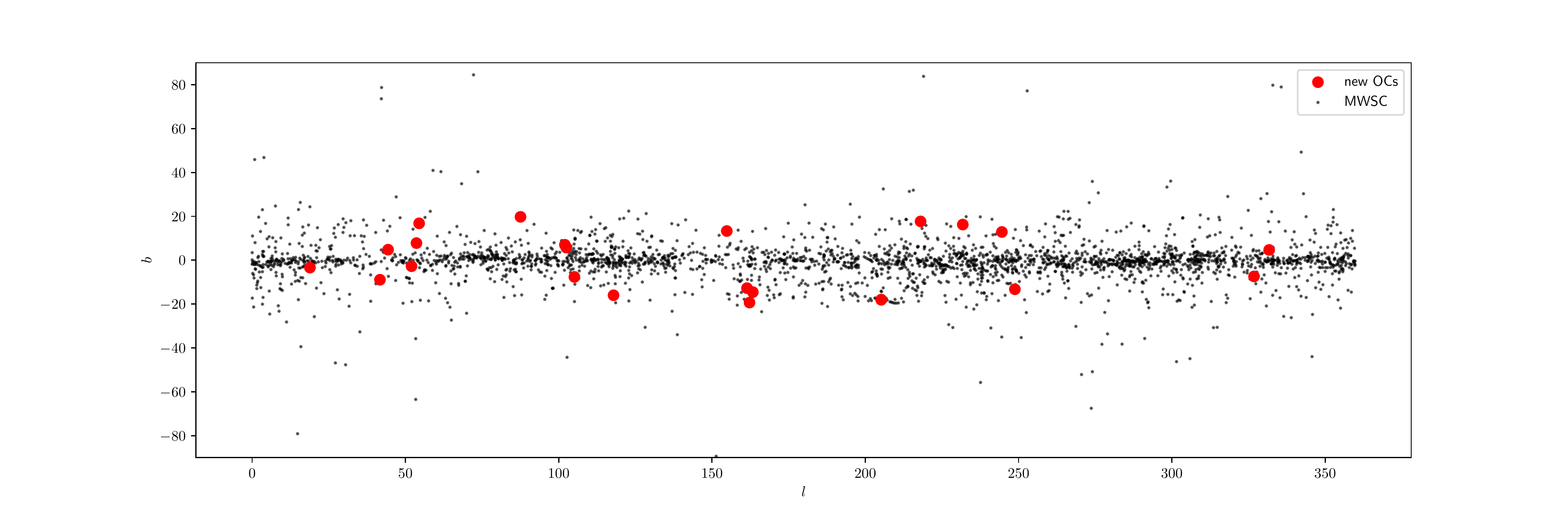}
\caption{Spatial distribution in $(l,b)$ of the found OCs (red) together with the ones listed in MWSC (black). The confirmed OCs tend to be at latitudes $|b| > 5\Unit{deg}$.}
\label{fig:spatial_dist}
\end{figure*}

The confirmed OCs are distributed on the Galactic disk, and they tend to be at galactic latitudes $|b| > 5\Unit{deg}$. \figrefalt{fig:spatial_dist} shows the distribution of the found OCs together with the ones listed in MWSC. They are also nearby compared to those in MWSC (see \figref{fig:par_dist}), most of them within $1\Unit{kpc}$ with the exception of UBC3, UBC6 and UBC27 which are detected with parallax of value $0.58 \pm 0.04\Unit{mas}$, $0.67 \pm 0.01\Unit{mas}$ and $0.88 \pm 0.03\Unit{mas}$, respectively.

\subsubsection{UBC1 and UBC12}

As mentioned in \secref{subsec:OCcand}, UBC1 and UBC12 are RSG4 and RSG3, respectively, in \cite{SRESBG-2016}.
They are located at about 330 and 430~pc, respectively. 
There is a rather good agreement in the proper motions of RSG3. On the contrary, for RSG4 the values are significant discrepant at the level of $12\sigma$.

\subsubsection{UBC3}

UBC3 is also a poor cluster located at about 1.7~kpc, the farthest cluster among our confirmed candidates. The presence of stars in the red clump area indicate an intermediate age cluster. There are only two stars with radial velocity in DR2 and both are in disagreement. One
of those stars is also discordant in terms of its position in the CMD. This could be indicative
of a non-membership.

\subsubsection{UBC4, UBC19 and UBC31}

UBC19 and UBC31 have proper motions and parallaxes compatible with being substructures
of the association Per~OB2, if we accept sizes of more than 8~deg for the association. If its is part or not of
Per~OB2 should be investigated through a deep study of a large area.
UBC19 has a celestial position near to Alessi Teustch~10 cluster in \cite{dias}, 
but their proper motions do not match. UBC4 has similar paramters but lies a bit farther at about 570~pc.

\subsubsection{UBC7 and Collinder 135}

\begin{figure}
\centering
\includegraphics[width=0.46\textwidth]{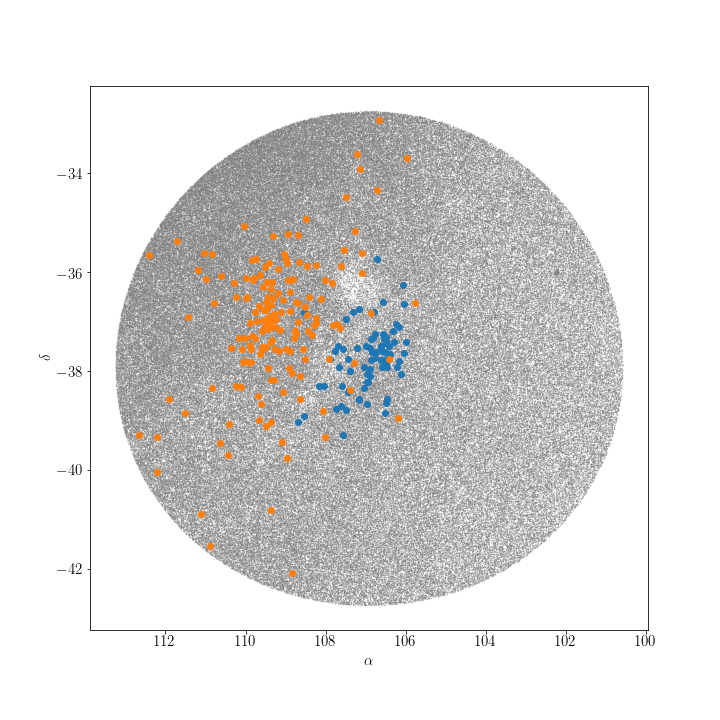}
\caption{Cone search of $5\Unit{deg}$ in the area of UBC7 (blue) and Collinder~135 (orange). The grey dots correspond to the stars brighter than $G=17$~mag with more than $120$ photometric observations in \gdrtwo data. We have checked that the lower stellar density between the two clusters only appears for parallaxes smaller than $1.5\Unit{mas}$, and so it is caused by dust in the background and does not impact our results for the clusters.}
\label{fig:UBC7_DR2}
\end{figure}

{\em Gaia} DR2 data allows to study UBC7 and Collinder~135 at fainter magnitudes than TGAS. The DR2 data do not show the scanning law pattern that TGAS shows, and still we see two concentrations on the sky (see Fig.~\ref{fig:UBC7_DR2}) with slightly different mean proper motions and parallaxes. The values of the mean and error of the mean for UBC7 are ($\mu_{\alpha^*}$,$\mu_{\delta}$)$=$($-9.74\pm 0.02$, $6.99\pm 0.02$) mas yr$^{-1}$ and $\varpi = 3.563\pm 0.006$ mas and for Collinder~135 are ($\mu_{\alpha^*}$,$\mu_{\delta}$)$=$($-10.09\pm 0.02$, $6.20\pm 0.03$) mas yr$^{-1}$ and $\varpi= 3.310\pm 0.004$ mas (computed with the members found with the method described in this paper). To discard possible artefacts due to regional systematics effects \citep{gdr2-Lindegren}, we have used the photometry and inspected the CMDs. The sequences overlap and thus reveal that both clusters have the same age or very similar. When apparent magnitudes are converted into absolute magnitudes using the individual parallaxes of the stars, the overlap of the two sequences is even better. This confirms that the difference in parallax is a true difference and not an artefact.

Therefore, given the differences in proper motions and parallaxes and given the separation in the sky, we conclude that they are two distinct groups, most probably formed in the same process given the similarity of their ages.

\subsubsection{UBC10}

This is a rather sparse cluster according to the members derived for the analysis in an
area of 1~deg radius with {\em Gaia} DR2. In addition, the celestial position and 
parallax of this cluster indicate a potential relationship with Cep~OB2 association. 
Therefore, we have explored a larger area of 2~deg and there are several subgroups of
proper motions and parallaxes certainly distributed towards the position of Cep~OB2.
A global analysis of an even larger area can confirm or discard the existence of new 
subgroups in this association. 

\subsubsection{UBC17}

The large sample of stars of {\em Gaia} DR2 with respect to TGAS has revealed two
groups of proper motions and parallaxes. The distances and proper motions relate them
to Ori~OB1 association. Exploring a larger area of $2\Unit{deg}$ we can identify ACCC19, Collinder 170 and sigma Ori clusters. This is an indication of the rich structure of the region and so a global analysis of an even larger area encompassing the whole Ori OB1 association is needed, but it goes beyond the goal of this paper.

\subsubsection{UBC32}

UBC20 TGAS DBSCAN candidate cluster was located at a parallax of about 0.5 mas. However, during the
analysis of {\em Gaia} DR2, such a cluster was not found but a clear detection at a parallax of 3.5 mas has been revealed. It is 
poor and sparse and decentered with respect to the studied area towards lower galactic 
latitudes.

\section{Conclusions}
\label{sec:conclusions}

We have designed, implemented and tested an automated data mining system for the detection of OCs using astrometric data. The method is based on i) DBSCAN, an unsupervised learning algorithm to find groups of stars in a $N$-dimensional space (our implementation uses five parameters $l$, $b$, $\parallax$, $\mu_{\alpha^*}$, $\mu_{\delta}$) and ii) an Artificial Neural Network trained to distinguish between real OCs and spurious statistical clusters by analysis of Color-Magnitude diagrams. It is designed to work with minimal manual intervention for its application to large datasets, and in particular to the \gaia Second Data Release \gdrtwo.

In this paper we have tuned and tested the performance of the method by running it using the simulated data and the TGAS dataset, which is small enough to manually check the results. This execution has generated a list of detections that after removal of know OCs from MWSC contains 31 new candidates. Using \gdrtwo data we manually examinated these candidates and confirmed around $70\%$ of them as open clusters, with a $100\%$ success in $N_{\rm found} > 5$. In addition, in the confirmation step, we are able to spot richer structures in particular regions that have to be further studied. 

From this exercise we have confirmed that our method can reliably detect OCs. We have also shown that the TGAS data contains some artefacts due to the nature of the \textit{Gaia} scanning law. We expect these effects to be much reduced (but not completely removed) in \gaia DR2, which includes the observations of 22 months of data and where the sky coverage is much more uniform (see \cite{gdr2-Lindegren}). Also, the bright limiting magnitude of TGAS prevented the detection of distant (and therefore faint) clusters, which will be detected with the much deeper \gdrtwo data. 

Finally, the method leads to reliable results but we have also identified some limitations. On the one hand, the representativeness of the training dataset for the ANN is crucial to distinguish real and non-real OCs, and we need to build a wider and more realistic training set of CMDs of OCs to use with \gaia DR2. On the other hand, since OCs look more compact or more sparse depending on their distance, there is not a universal value of the $\epsilon$ parameter in DBSCAN that can allow the detection of all of them. Therefore, this parameter needs to be adapted to the different possible characteristics of OCs in DR2.

\begin{acknowledgements}

This work has made use of results from the European Space Agency (ESA)
space mission {\it Gaia}, the data from which were processed by the {\it Gaia
Data Processing and Analysis Consortium} (DPAC).  Funding for the DPAC
has been provided by national institutions, in particular the
institutions participating in the {\it Gaia} Multilateral Agreement. The
{\it Gaia} mission website is \url{http: //www.cosmos.esa.int/gaia}. The
authors are current or past members of the ESA {\it Gaia} mission team and
of the {\it Gaia} DPAC.

This work was supported by the MINECO (Spanish Ministry of Economy) through 
grant ESP2016-80079-C2-1-R (MINECO/FEDER, UE) and ESP2014-55996-C2-1-R 
(MINECO/FEDER, UE) and MDM-2014-0369 of ICCUB (Unidad de Excelencia 'María de Maeztu'). 

This research has made use of the TOPCAT \citep{topcat}.
This research has made use of the VizieR catalogue access tool, CDS,
Strasbourg, France. The original description of the VizieR service was
published in A$\&$AS 143, 23
\end{acknowledgements}

\bibliographystyle{aa} 
\bibliography{bibliography}

\onecolumn
\begin{appendix}

\section{Color-Magnitude Diagrams of the found OCs}
  \label{sec:app_cmd}

\begin{figure*}[htb]
\centering
\includegraphics[width = 1.\textwidth]{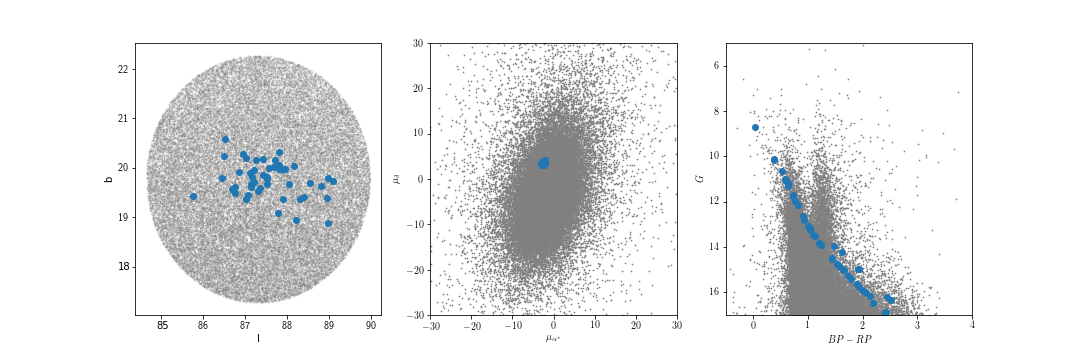}
\caption{Member stars (blue) together with field stars (grey) for UBC1 in $(l,b)$ (left) and in proper motion space (middle). The Color-Magnitude Diagram shows the sequence of the identified members (outlining an empirical isochrone) (right).}
\label{fig:cluster_summary_app_0}
\end{figure*}

\begin{figure*}[htb]
\centering
\includegraphics[width = 1.\textwidth]{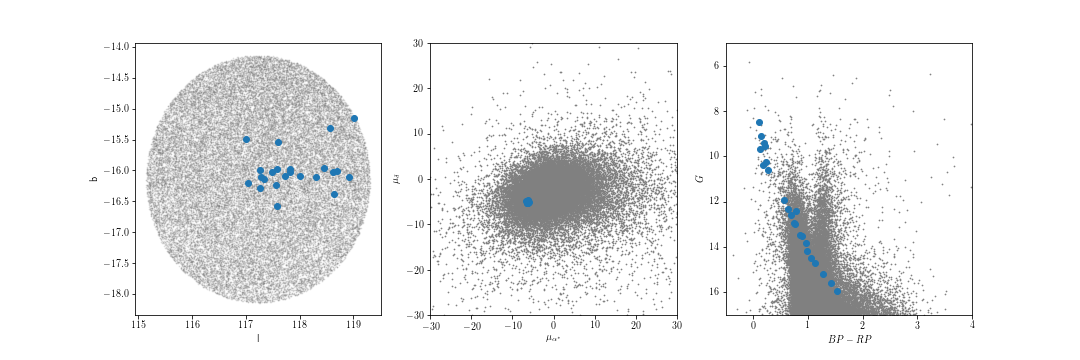}
\caption{Member stars (blue) together with field stars (grey) for UBC2 in $(l,b)$ (left) and in proper motion space (middle). The Color-Magnitude Diagram shows the sequence of the identified members (outlining an empirical isochrone) (right).}
\label{fig:cluster_summary_app_1}
\end{figure*}

\begin{figure*}[htb]
\centering
\includegraphics[width = 1.\textwidth]{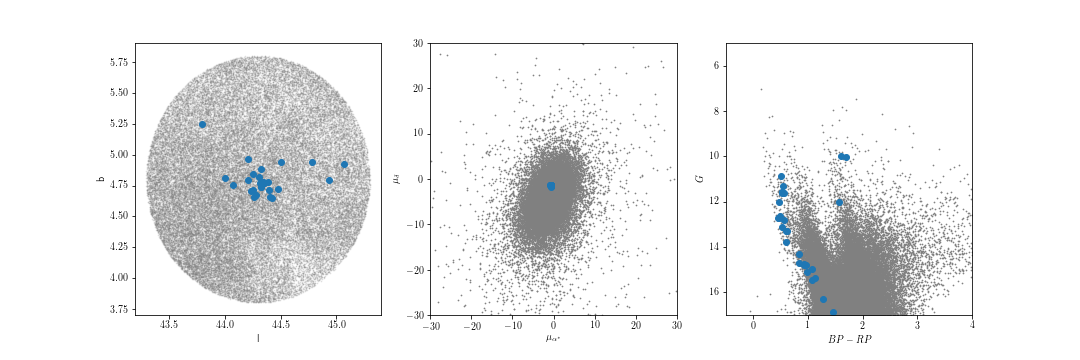}
\caption{Member stars (blue) together with field stars (grey) for UBC3 in $(l,b)$ (left) and in proper motion space (middle). The Color-Magnitude Diagram shows the sequence of the identified members (outlining an empirical isochrone) (right).}
\label{fig:cluster_summary_app_2}
\end{figure*}

\begin{figure*}[htb]
\centering
\includegraphics[width = 1.\textwidth]{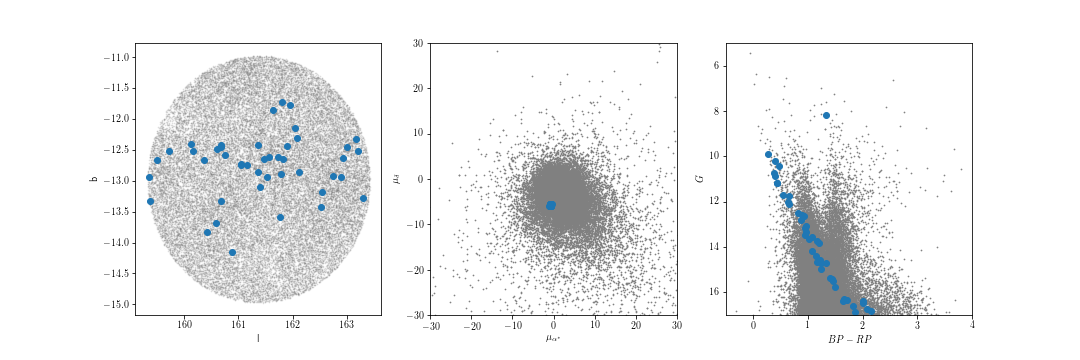}
\caption{Member stars (blue) together with field stars (grey) for UBC4 in $(l,b)$ (left) and in proper motion space (middle). The Color-Magnitude Diagram shows the sequence of the identified members (outlining an empirical isochrone) (right).}
\label{fig:cluster_summary_app_3}
\end{figure*}

\begin{figure*}[htb]
\centering
\includegraphics[width = 1.\textwidth]{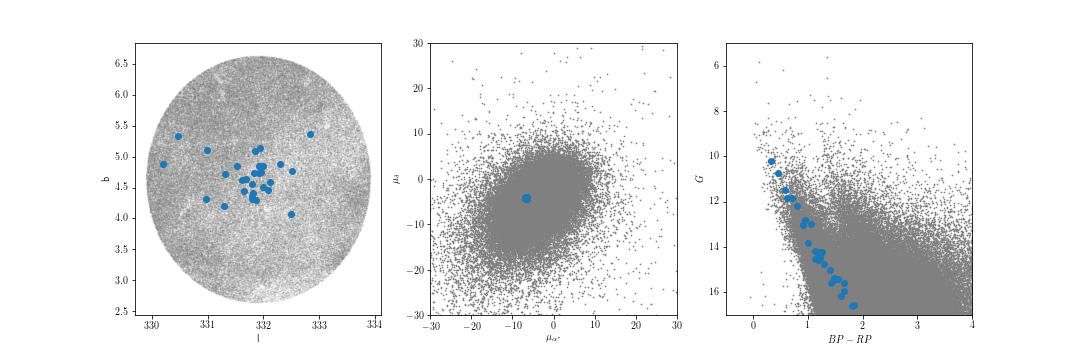}
\caption{Member stars (blue) together with field stars (grey) for UBC5 in $(l,b)$ (left) and in proper motion space (middle). The Color-Magnitude Diagram shows the sequence of the identified members (outlining an empirical isochrone) (right).}
\label{fig:cluster_summary_app_4}
\end{figure*}

\begin{figure*}[htb]
\centering
\includegraphics[width = 1.\textwidth]{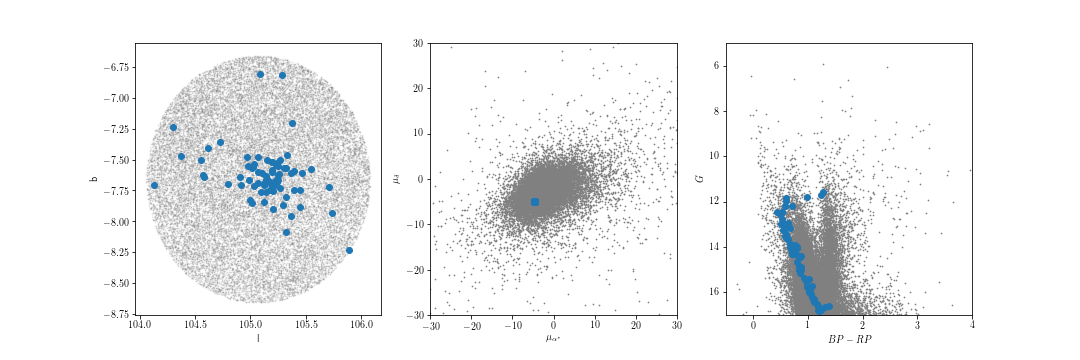}
\caption{Member stars (blue) together with field stars (grey) for UBC6 in $(l,b)$ (left) and in proper motion space (middle). The Color-Magnitude Diagram shows the sequence of the identified members (outlining an empirical isochrone) (right).}
\label{fig:cluster_summary_app_5}
\end{figure*}

\begin{figure*}[htb]
\centering
\includegraphics[width = 1.\textwidth]{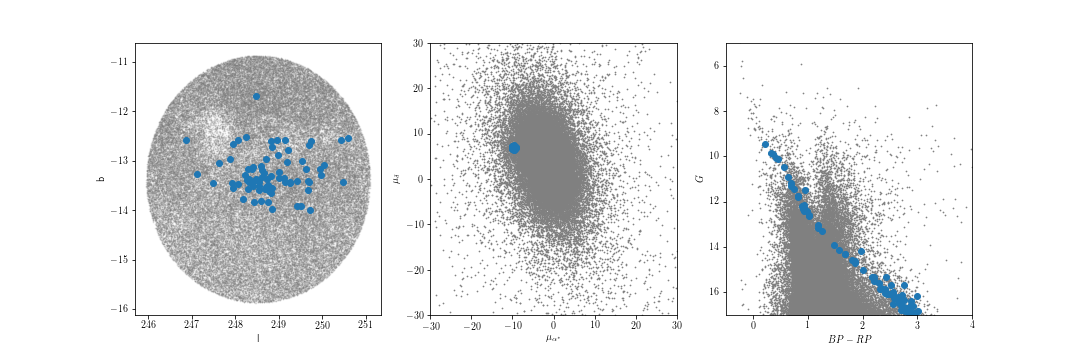}
\caption{Member stars (blue) together with field stars (grey) for UBC7 in $(l,b)$ (left) and in proper motion space (middle). The Color-Magnitude Diagram shows the sequence of the identified members (outlining an empirical isochrone) (right).}
\label{fig:cluster_summary_app_6}
\end{figure*}

\begin{figure*}[htb]
\centering
\includegraphics[width = 1.\textwidth]{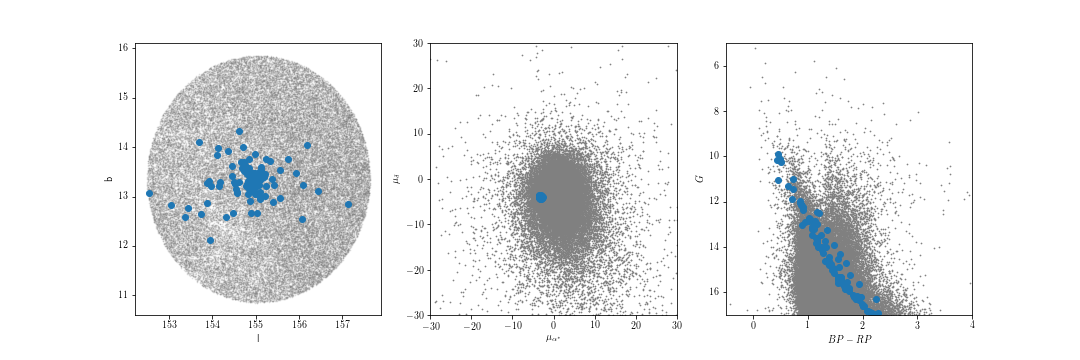}
\caption{Member stars (blue) together with field stars (grey) for UBC8 in $(l,b)$ (left) and in proper motion space (middle). The Color-Magnitude Diagram shows the sequence of the identified members (outlining an empirical isochrone) (right).}
\label{fig:cluster_summary_app_7}
\end{figure*}

\begin{figure*}[htb]
\centering
\includegraphics[width = 1.\textwidth]{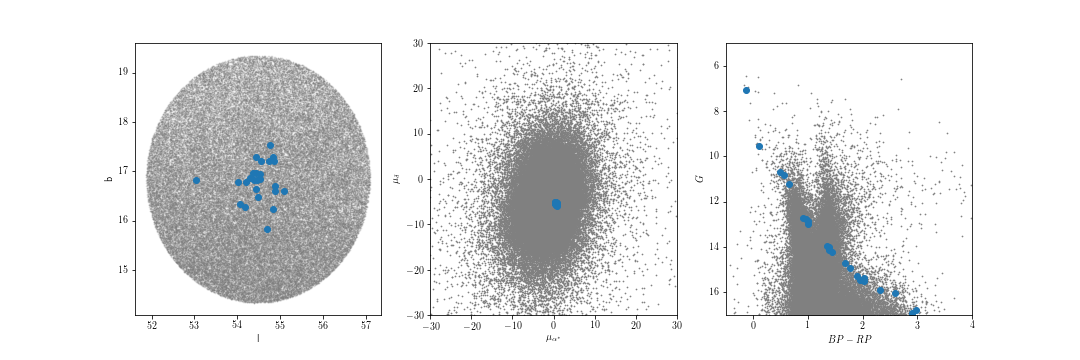}
\caption{Member stars (blue) together with field stars (grey) for UBC9 in $(l,b)$ (left) and in proper motion space (middle). The Color-Magnitude Diagram shows the sequence of the identified members (outlining an empirical isochrone) (right).}
\label{fig:cluster_summary_app_8}
\end{figure*}

\begin{figure*}[htb]
\centering
\includegraphics[width = 1.\textwidth]{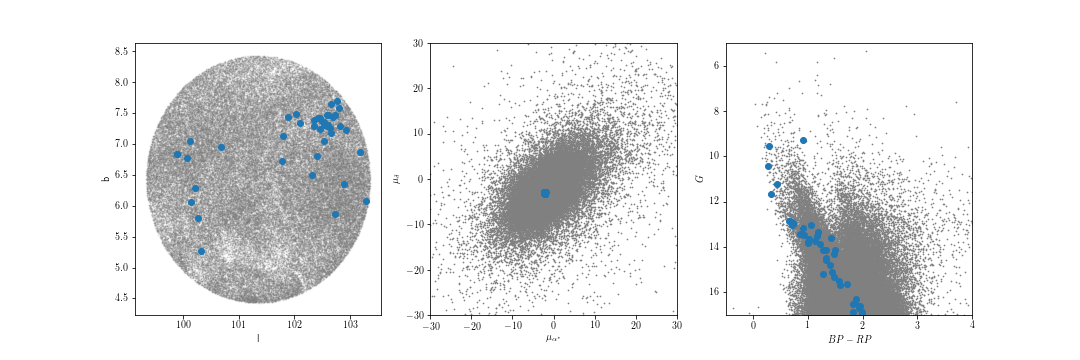}
\caption{Member stars (blue) together with field stars (grey) for UBC10a in $(l,b)$ (left) and in proper motion space (middle). The Color-Magnitude Diagram shows the sequence of the identified members (outlining an empirical isochrone) (right).}
\label{fig:cluster_summary_app_9a}
\end{figure*}

\begin{figure*}[htb]
\centering
\includegraphics[width = 1.\textwidth]{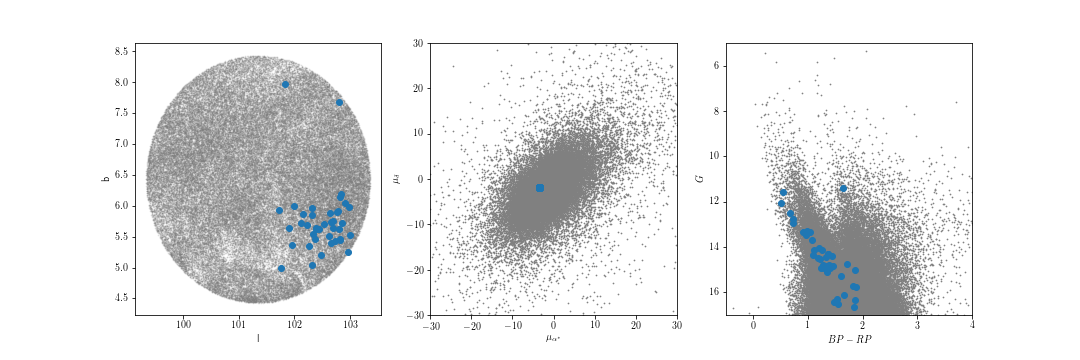}
\caption{Member stars (blue) together with field stars (grey) for UBC10b in $(l,b)$ (left) and in proper motion space (middle). The Color-Magnitude Diagram shows the sequence of the identified members (outlining an empirical isochrone) (right).}
\label{fig:cluster_summary_app_9b}
\end{figure*}

\begin{figure*}[htb]
\centering
\includegraphics[width = 1.\textwidth]{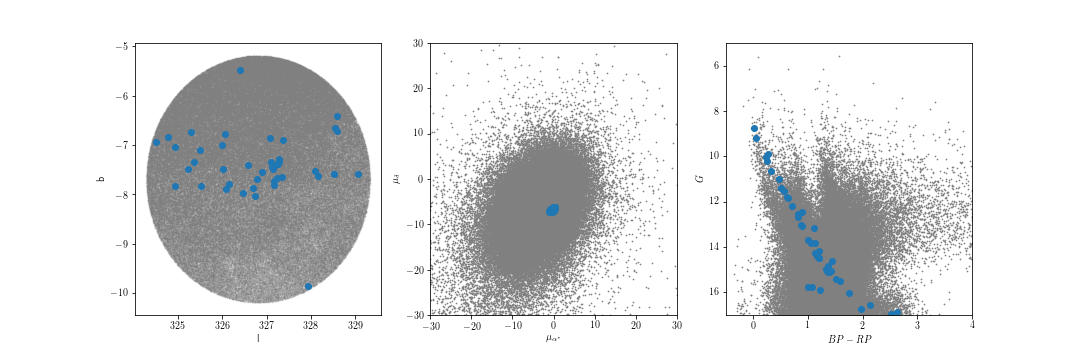}
\caption{Member stars (blue) together with field stars (grey) for UBC11 in $(l,b)$ (left) and in proper motion space (middle). The Color-Magnitude Diagram shows the sequence of the identified members (outlining an empirical isochrone) (right).}
\label{fig:cluster_summary_app_10}
\end{figure*}

\begin{figure*}[htb]
\centering
\includegraphics[width = 1.\textwidth]{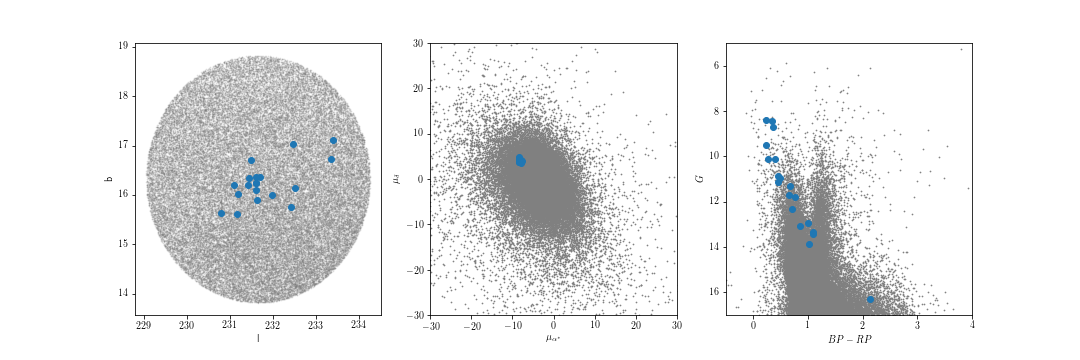}
\caption{Member stars (blue) together with field stars (grey) for UBC12 in $(l,b)$ (left) and in proper motion space (middle). The Color-Magnitude Diagram shows the sequence of the identified members (outlining an empirical isochrone) (right).}
\label{fig:cluster_summary_app_11}
\end{figure*}

\begin{figure*}[htb]
\centering
\includegraphics[width = 1.\textwidth]{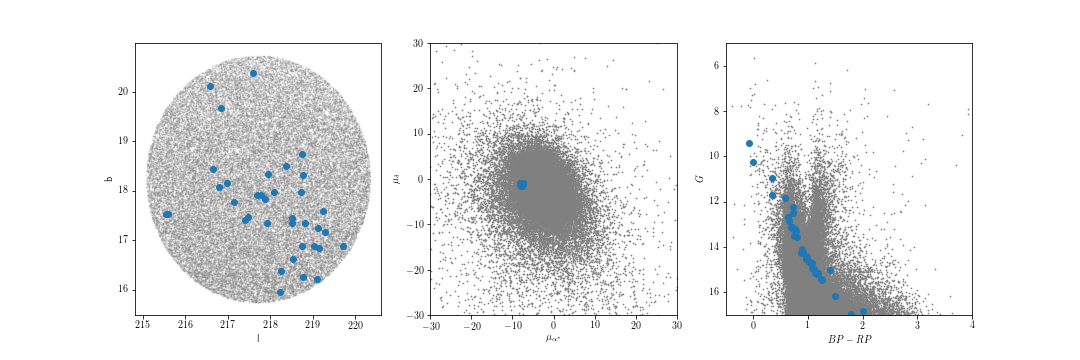}
\caption{Member stars (blue) together with field stars (grey) for UBC13 in $(l,b)$ (left) and in proper motion space (middle). The Color-Magnitude Diagram shows the sequence of the identified members (outlining an empirical isochrone) (right).}
\label{fig:cluster_summary_app_12}
\end{figure*}

\begin{figure*}[htb]
\centering
\includegraphics[width = 1.\textwidth]{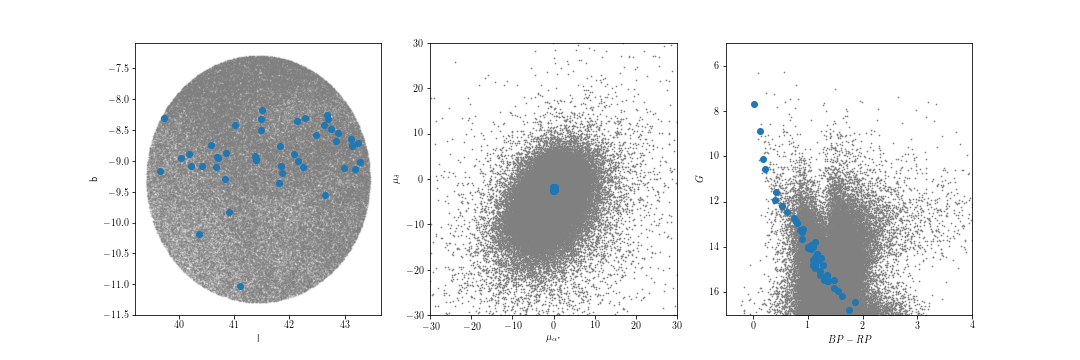}
\caption{Member stars (blue) together with field stars (grey) for UBC14 in $(l,b)$ (left) and in proper motion space (middle). The Color-Magnitude Diagram shows the sequence of the identified members (outlining an empirical isochrone) (right).}
\label{fig:cluster_summary_app_13}
\end{figure*}

\begin{figure*}[htb]
\centering
\includegraphics[width = 1.\textwidth]{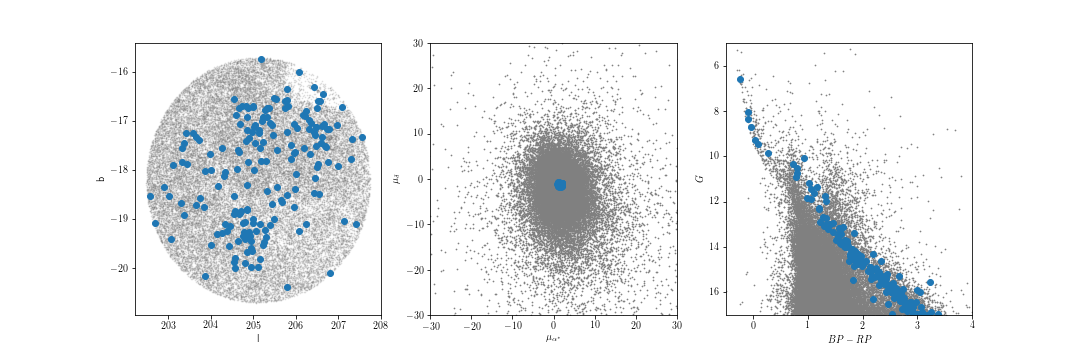}
\caption{Member stars (blue) together with field stars (grey) for UBC17a in $(l,b)$ (left) and in proper motion space (middle). The Color-Magnitude Diagram shows the sequence of the identified members (outlining an empirical isochrone) (right).}
\label{fig:cluster_summary_app_16a}
\end{figure*}

\begin{figure*}[htb]
\centering
\includegraphics[width = 1.\textwidth]{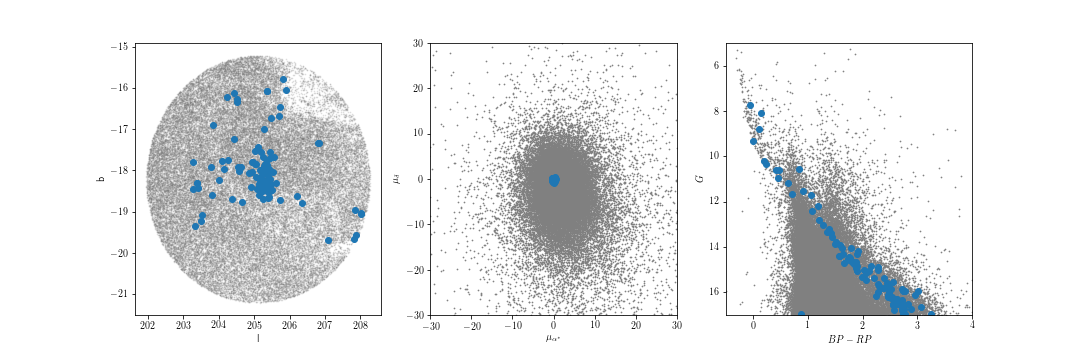}
\caption{Member stars (blue) together with field stars (grey) for UBC17b in $(l,b)$ (left) and in proper motion space (middle). The Color-Magnitude Diagram shows the sequence of the identified members (outlining an empirical isochrone) (right).}
\label{fig:cluster_summary_app_16b}
\end{figure*}

\begin{figure*}[htb]
\centering
\includegraphics[width = 1.\textwidth]{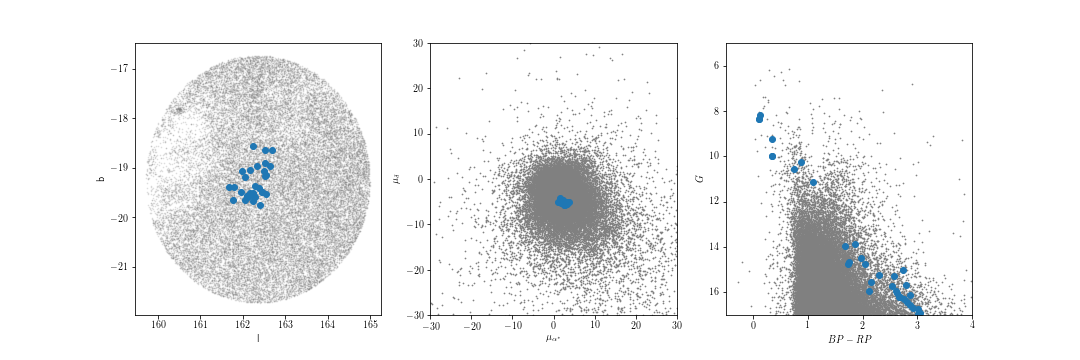}
\caption{Member stars (blue) together with field stars (grey) for UBC19 in $(l,b)$ (left) and in proper motion space (middle). The Color-Magnitude Diagram shows the sequence of the identified members (outlining an empirical isochrone) (right).}
\label{fig:cluster_summary_app_19}
\end{figure*}

\begin{figure*}[htb]
\centering
\includegraphics[width = 1.\textwidth]{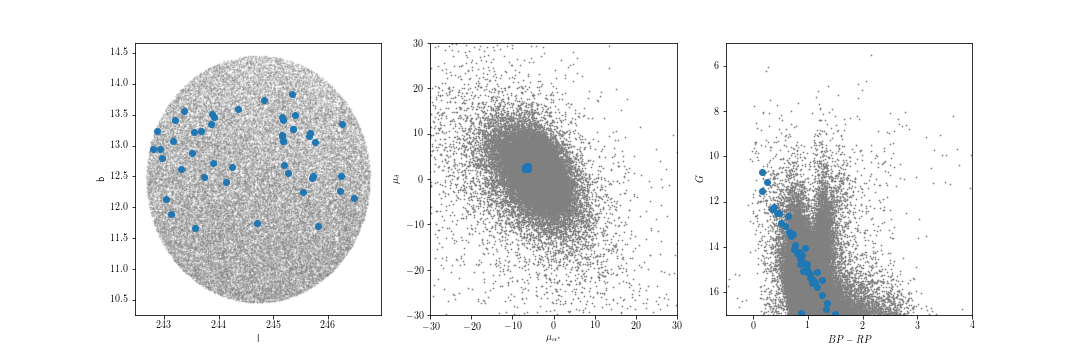}
\caption{Member stars (blue) together with field stars (grey) for UBC21 in $(l,b)$ (left) and in proper motion space (middle). The Color-Magnitude Diagram shows the sequence of the identified members (outlining an empirical isochrone) (right).}
\label{fig:cluster_summary_app_21}
\end{figure*}

\begin{figure*}[htb]
\centering
\includegraphics[width = 1.\textwidth]{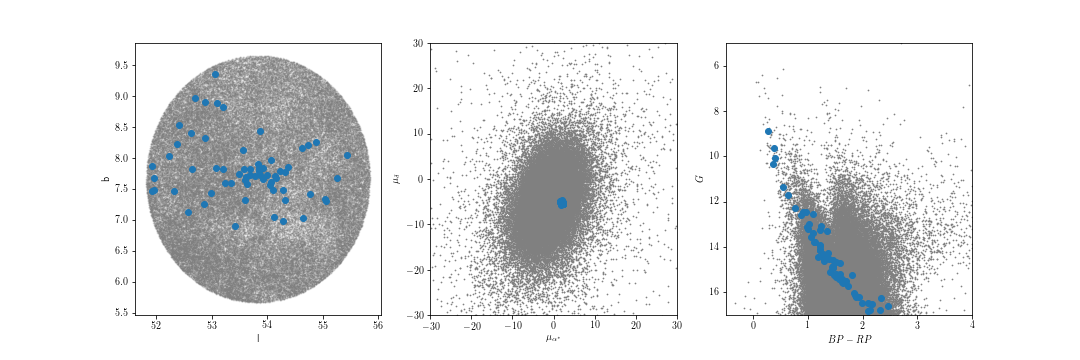}
\caption{Member stars (blue) together with field stars (grey) for UBC26 in $(l,b)$ (left) and in proper motion space (middle). The Color-Magnitude Diagram shows the sequence of the identified members (outlining an empirical isochrone) (right).}
\label{fig:cluster_summary_app_26}
\end{figure*}

\begin{figure*}[htb]
\centering
\includegraphics[width = 1.\textwidth]{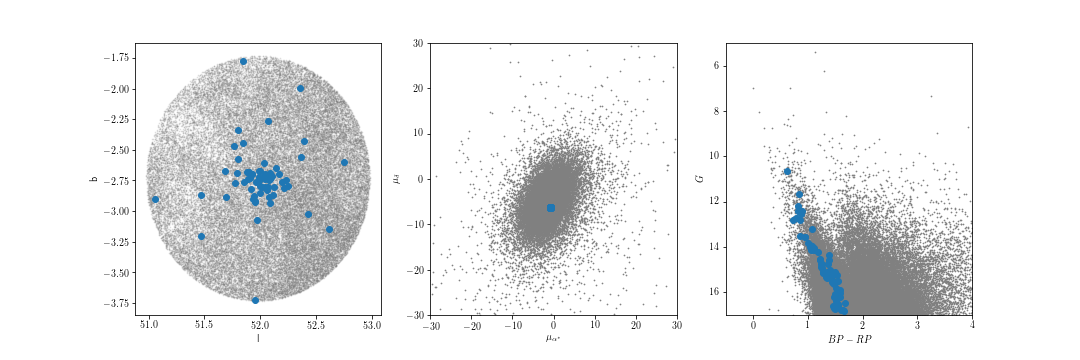}
\caption{Member stars (blue) together with field stars (grey) for UBC27 in $(l,b)$ (left) and in proper motion space (middle). The Color-Magnitude Diagram shows the sequence of the identified members (outlining an empirical isochrone) (right).}
\label{fig:cluster_summary_app_27}
\end{figure*}

\begin{figure*}[htb]
\centering
\includegraphics[width = 1.\textwidth]{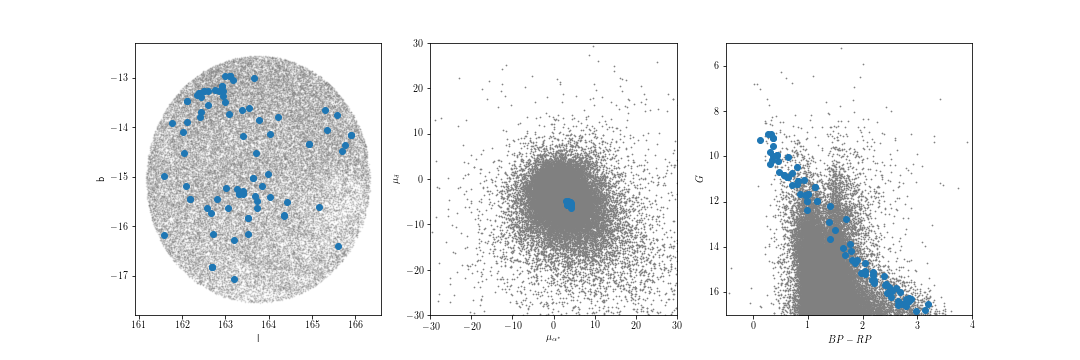}
\caption{Member stars (blue) together with field stars (grey) for UBC31 in $(l,b)$ (left) and in proper motion space (middle). The Color-Magnitude Diagram shows the sequence of the identified members (outlining an empirical isochrone) (right).}
\label{fig:cluster_summary_app_31}
\end{figure*}

\begin{figure*}[htb]
\centering
\includegraphics[width = 1.\textwidth]{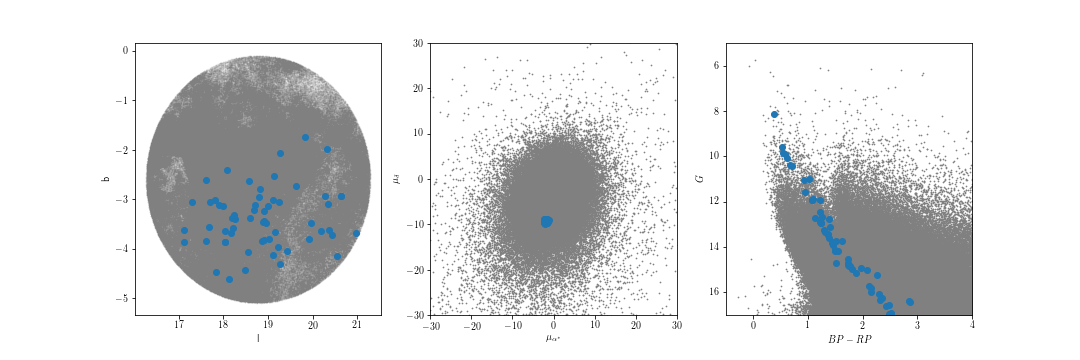}
\caption{Member stars (blue) together with field stars (grey) for UBC32 in $(l,b)$ (left) and in proper motion space (middle). The Color-Magnitude Diagram shows the sequence of the identified members (outlining an empirical isochrone) (right).}
\label{fig:cluster_summary_app_20}
\end{figure*}

\end{appendix}

\end{document}